\documentclass[%
 reprint,
 amsmath,amssymb,
 aps,
 superscriptaddress,
 prl,
 twocolumn,
 titling,
]{revtex4}

\usepackage{graphicx}%
\usepackage{dcolumn}%
\usepackage{bm}%
\usepackage{braket}
\usepackage{mathrsfs}

\usepackage{comment}

\begin{document}

\preprint{APS/123-QED}

\title{A geometric bound on the efficiency of irreversible thermodynamic cycles}%

\author{Adam G. Frim}%
\email{adamfrim@berkeley.edu}
\affiliation{%
 Department of Physics, University of California, Berkeley, Berkeley, California, 94720
}%
\author{Michael R. DeWeese}
\email{deweese@berkeley.edu}
\affiliation{%
 Department of Physics, University of California, Berkeley, Berkeley, California, 94720
}%
\affiliation{%
Redwood Center For Theoretical Neuroscience, University of California, Berkeley, Berkeley, California, 94720
}
\affiliation{%
Helen Wills Neuroscience Institute, University of California, Berkeley, Berkeley, California, 94720
}%

\date{\today}%

\begin{abstract}
Stochastic thermodynamics has revolutionized our understanding of heat engines operating in finite time. Recently, numerous studies have considered the optimal operation of thermodynamic cycles acting as heat engines with a given profile in thermodynamic space (e.g. $P-V$ space in classical thermodynamics), with a particular focus on the Carnot engine. In this work, we use the lens of thermodynamic geometry to explore the full space of thermodynamic 
cycles with continuously-varying bath temperature in search of optimally shaped cycles acting in the slow-driving regime. We apply classical  isoperimetric inequalities to derive a universal geometric bound on the efficiency of any irreversible thermodynamic cycle and explicitly construct efficient heat engines operating in finite time that nearly saturate this bound for a specific model system. Given the bound, these optimal cycles perform more efficiently than all other thermodynamic cycles operating as heat engines in finite time, including notable cycles, such as those of Carnot, Stirling, and Otto. For example, in comparison to recent experiments, this corresponds to orders of magnitude improvement in the efficiency of engines operating in certain time regimes. Our results suggest novel design principles for future mesoscopic heat engines and are ripe for experimental investigation.
\end{abstract}

\maketitle

\textit{Introduction}. Over the past several decades, stochastic thermodynamics has dramatically improved our understanding of nonequilibrium statistical physics \cite{1997_PRL_Jarzynski,Sekimoto,1999_PRE_Crooks,2005_PRL_Seifert,2010_PRL_Esposito_three,2010_PRL_Sagawa,2012_RPP_Seifert}. A major focus of study in this area has been the performance of engines operating in finite time, where both power and dissipation are finite, often with %
an emphasis on engines operating at maximal power \cite{1975_AJP_Curzon,2005_PRL_VandenBroeck,2007_EPL_Schmiedl,2009_PRL_Esposito,2010_PRL_Esposito,2015_PRX_Brandner,2016_PRL_Proesmans,2016_PRL_Shiraishi,2018_PRE_Ma_1,2018_PRE_Ma_2,2020_abiuso_prl,2020_PRL_Ma,2020_PRL_Miller,2021_PRE_Miangolarra_1,2022_PRE_Frim, 2022_PRR_Watanabe, 2022_PRXQ_alonso}. A recurring theme %
has been the interplay, %
and often incompatibility, among %
high efficiency, high output power, and low dissipation \cite{2015_PRX_Brandner,2016_PRL_Shiraishi,2020_abiuso_prl,2022_PRE_Frim}. To that end, we recently characterized optimal protocols for the finite-time operation of a Brownian Carnot engine \cite{2022_PRE_Frim}, a colloidal system introduced in \cite{2016_NatPhys_Martinez}, finding minimally dissipative cycles often came at the expense of reduced power and efficiency.

In this work, we study the implications of this interplay from a geometric perspective and arrive at a universal bound on the efficiency of heat engines operating in finite time. We also numerically construct optimal cycles that nearly saturate the bound, and we characterize its tightness for a specific model system and %
compare them to a variety of non-optimal cycles. %
We find that even natural extensions of well-known quasi-static cycles (e.g. Carnot engines) to finite-time cycles perform far less efficiently than the optimal cycles, %
demonstrating the utility of our result.

\textit{Efficiency of irreversible engines}.
Following ~\cite{2020_PRL_Brandner}, we use a definition of efficiency that directly captures the irreversibility of a thermodynamic cycle.
Specifically, for a generic thermodynamic engine operated by cyclically varying the temperature $T$ of a heat bath in contact with the system and some volume-like mechanical control variable $\lambda$, the (average) efficiency is defined as
\begin{equation}
    \label{eq:efficiency}
    \eta \equiv \frac{W}{U} = \frac{\oint X_\lambda d\lambda }{\oint X_T dT},
\end{equation}
where $X_\nu$ is the thermodynamic force conjugate to the control variable $\nu \in \{T,\lambda\}$, defined as 
\begin{align}
\label{eq:thermo_forces}
    &X_\lambda \equiv -\Braket{\frac{\partial H_{\bm{\Lambda}}}{\partial\lambda}},\\
    &X_T = S \equiv -\braket{\log \rho_t},
\end{align}
where $H_{\bm{\Lambda}}$ is the Hamiltonian of the (working) system for a fixed set $\bm{\Lambda} = (T,\lambda)$, $S$ is the system entropy, $\rho_t$ is the phase space distribution of the system at time $t$, and brackets denote ensemble averages. 
Note that $U$ here does not represent the internal energy, but can be thought of as the uptake of thermal energy from
the heat source, or the amount of energy that is available
for work production under a given temperature profile~\cite{2020_PRL_Brandner}.
This efficiency is well-defined for any engine with positive work output, %
$W>0$. Following the first law of thermodynamics and appealing to the cyclic operation of the engine, the efficiency may be rewritten as 
\begin{equation}
    \label{eq:efficiency_entropy}
    \eta = \frac{\oint X_\lambda d\lambda }{\oint S dT} =  \frac{\oint X_\lambda d\lambda }{ \oint X_\lambda d\lambda +\oint T d\Sigma}\leq 1,
\end{equation}
where $d\Sigma \equiv dS-dQ/T$ is the total infinitesimal entropy production in the universe. Here, $Q$ is the heat exchange into the system from the reservoir at a temperature $T$. Following the second law, $d\Sigma\geq 0$ such that the inequality follows directly and the unity bound can only be saturated for quasistatic, reversible engines. This definition of efficiency has gained traction in the study of finite-time heat engines \cite{2021_PRE_Miangolarra_1,2021_PRE_Miangolarra_2,2022_PRE_Frim}, and a comparable definition is standard for monothermal cycles, e.g. in active matter or chemical transduction contexts \cite{2016_JSM_Pietzonka,2019_PRX_Pietzonka,2020_PRE_Ekeh,2020_PRE_Lee}.

Now, let us consider a finite-time operation such that $\Sigma>0$ and the system is driven with finite driving rates $\dot{\bm{\Lambda}}$. The thermodynamic forces %
may be expanded as 
\begin{equation}
    \label{eq:lin_exp}
    X_\mu = \mathcal{X}_\mu - g_{\mu\nu}\dot{\Lambda}_\nu,
\end{equation}
where summation over repeated indices is implied. Here, $\mathcal{X}_\mu$ is the quasistatic value of $X$ for a given set $\bm{\Lambda}$ and, following the standard linear response framework, $g_{\mu\nu}$ is given by a correlation function
\begin{equation}
    \label{eq:metric}
    g_{\mu\nu} = \beta \int_0^\infty  dt \braket{\delta X_\mu(t)\delta X_\nu(0)}_{\bm{\Lambda}}.
\end{equation}
Using this expansion, the efficiency can be written

\begin{multline}
\eta = \frac{\oint X_\lambda d\lambda }{\oint X_TdT } \approx \frac{\oint (\mathcal{X}_\lambda-g_{\lambda\nu} \dot{\Lambda}_\nu  ) d\lambda}{\oint (\mathcal{X}_T-g_{T\nu} \dot{\Lambda}_\nu ) dT} \\
\approx 1-\frac{\oint dt \dot{\Lambda}_\mu g_{\mu\nu} \dot{\Lambda}_\nu }{\oint \mathcal{X}_\lambda d\lambda} = 1-\frac{A}{\mathcal{W}},
\end{multline}
where $A$ is the dissipated work (or alternatively called the dissipated availibility) through one %
cycle and $\mathcal{W}$ is 
the work output for quasistatic driving. Following \cite{1975_JCP_Weinhold,1979_PRA_Ruppeiner,1983_Salamon_PRL,1984_JCP_Salamon,1985_ZPB_Schlogl,1995_RMP_Ruppeiner,2007_PRL_Crooks,2012_PRL_Sivak}, %
we see that the thermodynamic control space is imbued with a geometric structure by reinterpreting $g_{\mu\nu}$ as a metric tensor. 
Geometric approaches such as this 
have greatly facilitated 
the development of optimal protocols for 
nonequilibrium systems \cite{2007_PRL_Crooks,2012_PRL_Sivak,2012_PRE_zulkowski,2014_PRE_Zulkowski_erasure,2013_PLOS_Zulkowski,2015_PRE_Zulkowski_quantum,2015_PRE_zulkowski_over,2015_PRE_Rotskoff,2020_abiuso_entropy,2020_JCP_Blaber}. 
With 
these definitions, the dissipated work satisfies
\begin{equation}
    \label{eq:A}
     \tau A \equiv \tau  \oint dt \dot{\Lambda}_\mu g_{\mu\nu} \dot{\Lambda}_\nu \geq \left(\oint dt \sqrt{\dot{\Lambda}_\mu g_{\mu\nu} \dot{\Lambda}_\nu}\right)^2 \equiv \mathcal{L}^2,
\end{equation}
where $\tau$ is the cycle duration and $\mathcal{L}$ is the \textit{thermodynamic length} of the protocol as defined by the metric. Optimal driving then yields~\cite{2020_PRL_Brandner}
\begin{equation}
    \label{eq:optimal_eta}
\eta \leq \eta^* = 1-\frac{\mathcal{L}^2}{\mathcal{W}\tau},
\end{equation}
which gives the optimal geometric efficiency $\eta^*$ for any closed curve in thermodynamic control space. 

The bound, Eq.~\eqref{eq:optimal_eta}, applies to any generic closed curve, yielding the best possible efficiency for an optimal temporal parametrization 
for a given path. We now seek to go further and find cycles that are maximally efficient over the space of all possible closed paths in thermodynamic control space. Fixing the protocol duration and given Eq.~\eqref{eq:optimal_eta}, to leading order in $1/\tau$, we must therefore find the curve that minimizes the quantity $\mathcal{L}^2/\mathcal{W}$. 

\textit{Isoperimetric bound on efficiency.}
We now note an important feature of $\mathcal{W}$. 
It is well known from standard thermodynamics %
that
the quasistatic work is simply given by the area contained within the curve of the (quasistatic) cycle in $\lambda - \mathcal{X}_\lambda$ space, which we will henceforth refer to as Clausius space in reference to the Clausius curve. We will refer to the inner region of this curve as $\mathcal{C}$ and the boundary
(the Clausius curve)
as $\partial \mathcal{C}$. By means of an appropriate change of coordinates, the metric $g_{\mu\nu}$ may be transformed for this space, yielding a metric $g_{\mu\nu}^{\mathcal{C}}$ such that the thermodynamic length is given
\begin{equation}
    \label{eq:L_clausius}
    \begin{split}
    \mathcal{L} &= \oint_{\partial \mathcal{C}} \sqrt{
    \begin{pmatrix} dT \\ d\lambda \end{pmatrix}_\mu
    g_{\mu\nu}
    \begin{pmatrix} dT \\ d\lambda \end{pmatrix}_\nu }\\
    &
    =\oint_{\partial \mathcal{C}} \sqrt{
    \begin{pmatrix} d\mathcal{X}_\lambda \\ d\lambda \end{pmatrix}_\mu
    g_{\mu\nu}^{\mathcal{C}}
    \begin{pmatrix} d\mathcal{X}_\lambda \\ d\lambda \end{pmatrix}_\nu }, 
    \end{split}
\end{equation}
and the quasistatic work is
\begin{equation}
    \label{eq:q_work}
    \mathcal{W} = \iint_{\mathcal{C}} d\lambda d\mathcal{X}_\lambda = \oint_{\partial \mathcal{C}} \mathcal{X}_\lambda d\lambda .
\end{equation}
We define one final quantity of interest: the thermodynamic area, defined as the integral over of the region with the proper thermodynamic geometric measure:
\begin{equation}
    \label{eq:thermo_area}
    \mathcal{A} = \iint_{\mathcal{C}} \sqrt{g^{\mathcal{C}}} d\lambda d\mathcal{X}_\lambda,
\end{equation}
where $g^{\mathcal{C}} \equiv \det{g_{\mu\nu}^{\mathcal{C}}}$. 
The physical interpretation of this quantity may not be as clear as that of thermodynamic length, though we will %
use it as a key ingredient. Notably, dating to antiquity \cite{payne_isoperimetric_1967,Osserman_1978}, bounds exist relating the perimeter $\mathscr{L}(D)$ and area $\mathscr{A}(D)$ of a region $D$, known as \textit{isoperimetric inequalities}. In general, for a simply-connected two-dimensional region, one has two inequalities \cite{Osserman_1978}
\begin{align}
    \label{eq:isoperimetric_eq_1}
    &\mathscr{L}^2 \geq 4\pi \mathscr{A}-2\left[\iint_D K^+ \right]\mathscr{A},\\
    \label{eq:isoperimetric_eq_2}
    &\mathscr{L}^2 \geq 4\pi \mathscr{A}-\bigg[\text{sup}_D K \bigg]\mathscr{A}^2.
\end{align}
where $K$ is the Gaussian curvature of the underlying space and $K^+(p) \equiv \text{max}(K(p),0)$ for a point $p\in D$. As a simple example, we remind the reader of the case of Euclidean space where $K = K^+ = 0$ and both inequalities yield $\mathscr{L}^2 \geq 4\pi \mathscr{A}$, which is saturated only for the optimal shape of a circle. Similar isoperimetrically optimal shapes exist for other manifolds \cite{Osserman_1978}. Importantly, whenever the curvature is everywhere nonpositive, Eq.~\eqref{eq:isoperimetric_eq_1} yields the familiar Euclidean bound, though it is only tight for special spaces, such as when $K=0$.

In the context of thermodynamic geometry, the bound readily applies, where now we consider the thermodynamic length of a closed curve and the thermodynamic area of the enclosed region. That is 
\begin{align}
    \label{eq:isoperimetric_bound_1}
    &\mathcal{L}^2 \geq 4\pi \mathcal{A}-2\left[\iint_{\mathcal{C}} K^+ \right]\mathcal{A},\\
    \label{eq:isoperimetric_bound_2}
    &\mathcal{L}^2 \geq 4\pi \mathcal{A}-\bigg[\text{sup}_{\mathcal{C}} K \bigg]\mathcal{A}^2.
\end{align}
Henceforth, we will focus on the case when $K\leq 0$ and will relegate the more general case to the Supplemental Material \cite{SM}. In this case, we may assume $\mathcal{L}^2\geq 4\pi \mathcal{A}$. Therefore,
\begin{equation}
    \label{eq:bound}
    \frac{\mathcal{L}^2}{\mathcal{W} }= \frac{\mathcal{L}^2}{\mathcal{W}} \frac{\mathcal{A}}{\mathcal{A}} = \frac{\mathcal{L}^2}{\mathcal{A}} \frac{\mathcal{A}}{\mathcal{W}} \geq 4\pi \frac{\mathcal{A}}{\mathcal{W}}
    =4\pi \overline{\sqrt{g^{\mathcal{C}}}},
\end{equation}
where the overline indicates an area average value. Given this bound, we may now write for the efficiency a universal bound 
\begin{equation}
    \label{geo_bound}
    \eta \leq 1-\frac{\mathcal{L}^2}{\mathcal{W}\tau} \leq 1-\frac{4\pi \overline{\sqrt{g^{\mathcal{C}}}}}{\tau}.
\end{equation}
This bound, which we will refer to as an isoperimentric bound on efficiency, is our first major result and presents a fundamentally new and purely geometric bound on the efficiency of nonequilibrium engines. Also, whereas the previous bound, Eq. \ref{eq:optimal_eta} was directly applicable to the optimized parametrization of a specific predetermined cycle, this bound places constraints on the optimal \textit{shape} of a cycle and can only be 
approached for geometrically %
optimized cycle shapes. This therefore introduces a new optimization principle wherein optimally efficient cycles must minimize the average value of $\sqrt{g^{\mathcal{C}}}$ over the region they enclose in Clausius space. In a pioneering study, a similar bound was recently recovered for the specific case of the  Brownian gyrator~\cite{2021_PRE_Miangolarra_2}, though we highlight that Eq.~\eqref{geo_bound} applies generically.

\textit{Construction of optimal cycles.}
Given this isoperimetric bound, we now seek to construct optimal cycles that (nearly) saturate it. This is done by finding shapes that minimize the ratio $\mathcal{L}^2/\mathcal{A}$. Inspired by previous literature \cite{Osserman_1978}, we will use a variational principle. Namely, we seek to maximize $\mathcal{A}$ while holding $\mathcal{L}^2$ fixed; we do so by means of a Lagrange multiplier. The relevant functional takes the form

\begin{equation}
    \label{eq:variational}
    \mathcal{F} = \oint dt \left(\sqrt{g^\mathcal{C}}X_\lambda \dot{\lambda} - \xi \begin{pmatrix} \dot{X}_\lambda \\ \dot{\lambda} \end{pmatrix}_\mu
    g_{\mu\nu}^{\mathcal{C}}
    \begin{pmatrix} \dot{X}_\lambda \\ \dot{\lambda} \end{pmatrix}_\nu  \right),
\end{equation}
where $\xi$ serves as a Lagrange multiplier enforcing a fixed dissipation for shapes that maximize the corresponding area \footnote{The dissipation $\oint (dX_\lambda , d\lambda) \cdot \mathbf{g} \cdot (dX_\lambda , d\lambda)$ is geometrically denoted the energy of the curve and optimization with respect to its value held fixed still optimizes the ratio $\mathcal{L}^2/\mathcal{A}$ while simultaneously outputting an optimal, arclength parametrization of the resultant curve.}. Optimization is found through solving the resulting Euler-Lagrange equations for $X_\lambda(t)$ and $\lambda(t)$ under a cyclic constraint. 

\textit{Parametric harmonic oscillator}
We will now 
illustrate the utility of this result by studying the 
parametric harmonic oscillator. %
This important model system consists of a particle of mass $m$ trapped in a harmonic potential with variable stiffness $V(x) = 1/2k(t)x^2$ in contact with a heat bath of variable temperature $T(t)$ and subject to viscous damping $\zeta$. This system has been %
studied extensively~\cite{2007_PRL_Schmiedl,2011_NatPhys_Blickle,2012_PRE_zulkowski,2015_PRL_Martinez,2016_NatPhys_Martinez,2017_SM_Martinez,2020_Huang,2022_PRE_Frim}, and its geometry has been well characterized --- %
the metric is %
\[g_{\mu\nu} = \frac{mk_B}{4\zeta} \begin{pmatrix}
\frac{1}{T} \left(4+\frac{(\zeta)^2}{km}\right) &
-\frac{1}{ k } \left(2+\frac{(\zeta)^2}{km}\right) \\
-\frac{1}{ k } \left(2+\frac{(\zeta)^2}{km}\right) &
\frac{T}{k^2} \left(1+\frac{(\zeta)^2}{km}\right) 
\end{pmatrix}_{\mu\nu}.\]
In this case, the mechanical control variable is $k$ and the corresponding force is $X_k = -1/2 \braket{x^2}$. In the quasistatic limit, we have $\mathcal{X}_k = -k_BT/2k$ by the equipartition theorem. Therefore, under the change of variables $(T,k) \rightarrow (\tilde{P},\tilde{V})\equiv  (\mathcal{X}_k,k)$, we find 
\[g_{\mu\nu}^{\mathcal{C}} =-\begin{pmatrix} \frac{\zeta}{2\tilde{P}} + \frac{2m\tilde{V}}{\zeta \tilde{P}} & \frac{m}{\zeta} \\
\frac{m}{\zeta} &
\frac{m\tilde{P}}{2\zeta \tilde{V}}
\end{pmatrix},
\]
where we use $\tilde{P}$ and $\tilde{V}$ to suggestively map to pseudo-pressure and pseudo-volume in Clausius space and recapture the usual identity $d\mathcal{W} = \tilde{P}d\tilde{V}$. 
We find $g^{\mathcal{C}} = m/(4\tilde{V})$ and the Gaussian curvature $K = 1/(\zeta \tilde{P}) \leq 0$ as $\tilde{P}\leq 0$ such that Eq.~\eqref{geo_bound} applies. As a result, we have
\[\eta \leq 1-\frac{2\pi \overline{\sqrt{\frac{m}{\tilde{V}}}}}{\tau}
\]
for any possible %
thermodynamic cycle. 

We now seek to characterize the optimal cycles for this model system. Following Eq.~\eqref{eq:variational}, we consider the functional
\begin{equation}
\label{eq:iso_functional}
    \mathcal{F} = \oint dt \left[\sqrt{g^\mathcal{C}} \tilde{P}\dot{\tilde{V}} + \xi \ \begin{pmatrix} \dot{\tilde{P}} \\ \dot{\tilde{V}} \end{pmatrix}
    \begin{pmatrix} \frac{\zeta}{2\tilde{P}} + \frac{2m\tilde{V}}{\zeta \tilde{P}} & \frac{m}{\zeta} \\
\frac{m}{\zeta} &
\frac{m\tilde{P}}{2\zeta \tilde{V}}
\end{pmatrix}
    \begin{pmatrix} \dot{\tilde{P}} \\ \dot{\tilde{V}} \end{pmatrix} \right].
\end{equation}
The Euler-Lagrange equations are then found by varying with respect to $\tilde{P}$ and $\tilde{V}$ (Supplemental Material~\cite{SM}). 
To our knowledge, 
the resulting Euler-Lagrange equations are analytically intractable, so we turn to numerics. %
For simplicity, we generate cycles from a given set of initial conditions, $\tilde{P}(0), \tilde{V}(0), \dot{\tilde{P}}(0),$ and $\dot{\tilde{V}}(0)$, and %
a given value of the Lagrange multiplier $\xi$. If we concern ourselves only with the shape of the curve rather than its particular parameterization, we could instead parameterize $\tilde{P}$ as a function of $\tilde{V}$, such that one of these initial condition degrees of freedom is redundant. %
Thus, optimal shapes are specified by a four-parameter family determined by $\tilde{P}(t=0), \tilde{V}(t=0), d\tilde{P}/d\tilde{V}(t=0)$ and $\xi$.

\begin{figure}[t]
    \centering
    \includegraphics[width=\columnwidth]{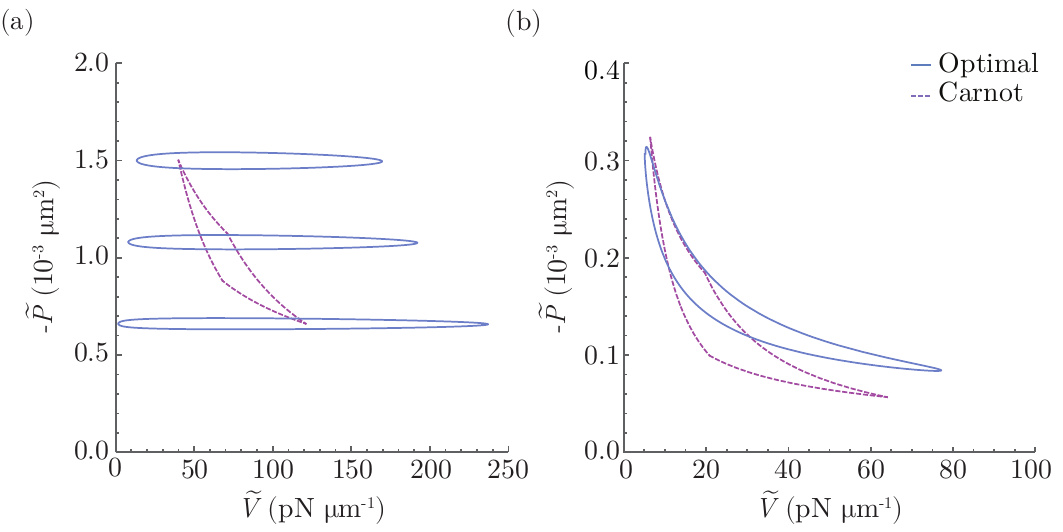}
    \caption{\textbf{(a)} A Brownian Carnot cycle (dashed purple) and three isoperimetrically optimal cycles (blue) with identical values of $\mathcal{W}$ and $\overline{\sqrt{g^\mathcal{C}}}$ using mass and damping values from \cite{2016_NatPhys_Martinez}. %
    The optimal cycles have a $~55\times$ increase in performance relative to the Carnot cycle. \textbf{(b)} Same as (a), %
    but in a more underdamped regime.}
     \label{fig:fig1}
\end{figure}

\begin{figure}[t]
    \centering
    \includegraphics[width=\columnwidth]{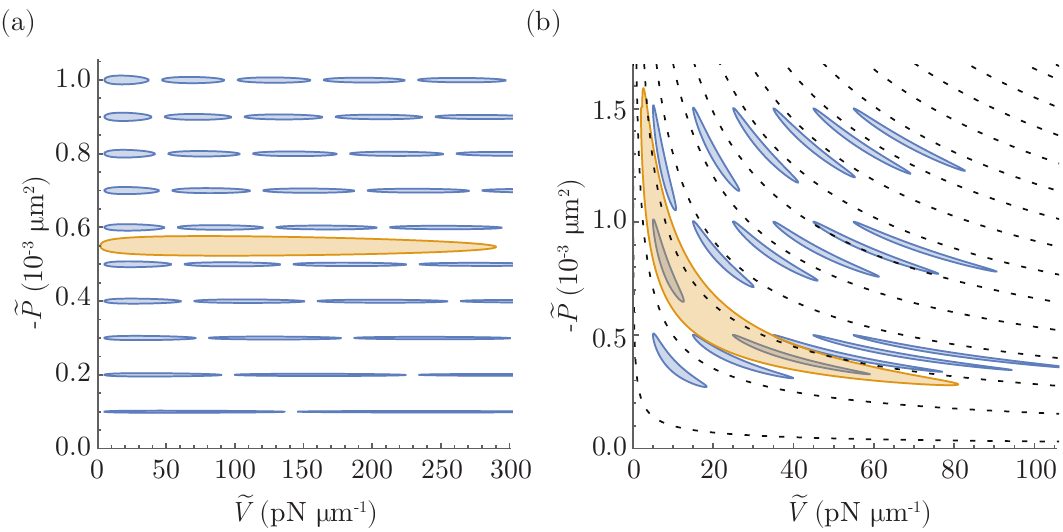}
    \caption{\text{\textbf{(a)}} Optimally constructed cycles with quasistatic output work of $6$ pN $\mu$m (blue) and 120 pN $\mu$m (orange) for experimental parameters used in \cite{2016_NatPhys_Martinez}. \textbf{(b)} Same as (a) for underdamped parameters; adiabats shown in dotted lines. }
    \label{fig:fig2}
\end{figure}

Generically, it is unclear whether %
these conditions will be sufficient to specify a smooth closed curve~%
\cite{payne_isoperimetric_1967,Osserman_1978,Ritore}.  In general, the isoperimetrically optimal curves will consist of stable smooth curves that have constant geodesic curvature at (nearly) all points \cite{payne_isoperimetric_1967,Osserman_1978,Ritore}.  In our case, the numerical solutions yield curves of constant geodesic curvature that are typically non-closed but instead consist of ``near-miss" cycloids~\cite{SM}, which has been observed previously %
while seeking optimal cycles~\cite{2020_Huang}. However, %
as we are interested only in cyclic engines, we will construct cycles that traverse a single optimal period by truncating the curve when it is at the nearest point on the curve whose tangent curve is parallel to the original tangent curve. We then connect these two points by a straight line, thus closing the cycle. For small cycles, the resulting %
kink is imperceptible whereas it becomes more noticeable for larger cycles, as can be seen in Figs.~\ref{fig:fig1} and~\ref{fig:fig2}. Similarly, the efficiencies are impacted more significantly for larger deviations from smooth curves. %
Truly optimal curves would consist of fully smooth closed curves of constant geodesic curvature, %
which may only be realizable in specific regions of the Clausius space. By choosing to work with specific initial conditions, we allow for the construction of near optimal cycles at all points of Clausius space at the cost of having to introduce small, finite sections of non-optimality. %
As we will see, these ``optimal" cycles still prove remarkably close to saturating the bound and strongly outperform all other cycles we consider.

For example, 
consider the Brownian Carnot cycle, whose shape is given in Fig.~\ref{fig:fig1}.
Setting $\zeta = 7.51 $ $\mu$g s$^{-1}$ and $m=0.545 $ pg (based on experimental parameters used in \cite{2016_NatPhys_Martinez}), and choosing the extremal values of $\tilde{P}$ and $\tilde{V}$ shown, 
Eq.~\eqref{eq:bound} implies that
its %
greatest possible efficiency 
is 
\[(1-\eta)\tau = \frac{\mathcal{L}^2}{\mathcal{W}} = 9.58\times 10^{-4} \text{ s} \gg 1.76\times 10^{-5} \text{ s} = \frac{4\pi \mathcal{A}}{\mathcal{W}},\]
demonstrating that the performance of the Carnot cycle is 
a factor of $55$ times larger than the optimal value for these parameter settings.
In contrast, by constructing a series of (nearly) optimally-shaped cycles with identical values of $\mathcal{W}$ and $\overline{\sqrt{g^{\mathcal{C}}}}$, we instead find the bound is %
approached to within three-hundredths of one percent. The shapes of these cycles are displayed in Fig.~\ref{fig:fig1}(a). 

We next consider a slightly underdamped parameter regime, setting $m=80.9$ mg and $\zeta=15.0$ mg s$^{-1}$. In this regime, various Carnot engine cycles (for 
given limits on hot and cold bath temperature) 
perform somewhat better, 
but they still do not come close to saturating %
it. 
Intriguingly, in contrast to the previous regime, optimally constructed shapes now appear much closer to Carnot-like cycles than those previously constructed, as in Fig~\ref{fig:fig1}(b). Now, we find that %
optimal engines seemingly remain close to adiabats for a significant duration of the cycle, %
albeit with a rapid (and smooth) turnaround at corners. This is replicated for other choices of maximal temperature and stiffness as shown in Fig.~\ref{fig:fig1}(b). We also show optimal cycles for a variety of values of $m/\zeta$ in the Supplemental Material \cite{SM}.
\begin{figure}[b]
    \centering
    \includegraphics[width=\linewidth]{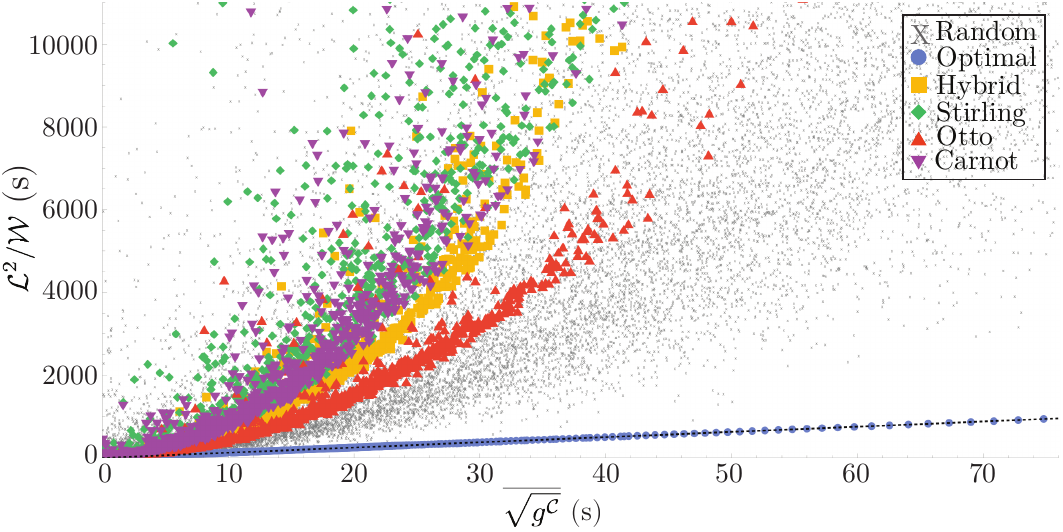}
    \caption{Optimally constructed cycles strongly outperform Carnot and all other cycles tested. Optimal (blue circles), Hybrid (orange squares), Stirling (green diamonds), Otto (red upwards triangle), Carnot (purple downwards triangle), and randomly constructed (gray X's) plotted against the bound Eq.~\eqref{eq:bound} (black dotted line).}
    \label{fig:fig3}
\end{figure}

We can further evaluate our bound by comparing it to generic cycles in Clausius space. In particular, in Fig.~\ref{fig:fig3}, we plot the value of $\overline{\sqrt{g^{\mathcal{C}}}}$ against $\mathcal{L}^2/\mathcal{W}$ for 15000 randomly constructed cycles, as well as 1000 randomly selected Carnot, Stirling, Otto, and hybrid Carnot/geodesic (introduced in~\cite{2022_PRE_Frim}) cycles (see \cite{SM} for a review of these various cycles and for details of how Fig.~3 is constructed). We also plot 500 optimally constructed cycles. As is easily observed, the optimal cycles nearly saturate the bound for a large parameter regime of $\overline{\sqrt{g^{\mathcal{C}}}}$ whereas other cycles %
are far less efficient. 
The data 
are surprisingly structured: %
for a fixed shape, we find $\mathcal{L}^2/\mathcal{W} \sim \big(\overline{\sqrt{g^{\mathcal{C}}}}\big)^2$, an interesting empirical finding. 
Cycles that saturate the bound must therefore \emph{not} maintain the same cyclical shape for different average values of $\sqrt{g^{\mathcal{C}}}$, but instead change so as to remain optimal, as was shown in Fig.~\ref{fig:fig2}.

\textit{Discussion.}
In this work, we have applied the classical isoperimetric inequalities to thermodynamic spaces, yielding a novel universal bound on the efficiency of any closed thermodynamic cycle for a generic system. We have then constructed optimal cycles that nearly saturate this bound for the specific model system of the parametric harmonic oscillator. Importantly, this bound, although not always tight, nevertheless strongly constrains the efficiency of thermodynamic cycles and introduces new design principles for the design of efficient finite-time heat engines. We emphasize that all results are independent of temporal parametrization and depend only on the shape of cycles.

This derivation ultimately arises from thermodynamic geometry, which is a perturbative solution to first order in response theory. Our bound is therefore only approximate in nature and does not technically constrain the efficiency of engines acting beyond slow-driving. However, it is unlikely such engines would prove less dissipative and more efficient than their slower-driven counterparts, such that one would expect the bound to still apply, but not be as tight.
A more interesting question is whether there are higher order or even nonperturbative bounds that constrain the efficiency of all finite time cycles.
Such bounds have been found for the fluctuations of the more traditional definition of efficiency about its mean~ \cite{2014_PRE_Verley, 2019_arxiv_Ito, 2021_PRL_Saryal}. Similarly, one should ask if the efficiency of optimal engines in our framework would still outperform others, such as the Carnot or Stirling engine, even when far away from linear response where our results are not guaranteed to apply.
We leave %
the detailed study of the question
to future work,
but there %
does exist anecdotal evidence that engines designed in linear response %
prove more efficient, even when operating far-from-equilibrium~\cite{2020_PRL_Brandner,2022_PRE_Frim}.

Also, although the main focus of this work was in optimizing the average efficiency of thermodynamic cycles, given that our optimal cycles were designed to be minimally dissipative for a fixed thermodynamic area, they should also have a high output power, though they may not be \textit{optimally} powerful cycles. A recurring theme in the literature is the set of tradeoffs between high output power, high efficiency, low dissipation, and minimal fluctuations about the means of stochastic thermodynamic quantities  \cite{2018_PRE_Ma_2, 2018_PRL_Pietzonka, 2020_abiuso_entropy,2020_PRL_Brandner,2020_PRL_Miller,2022_PRXQ_alonso,2022_PRR_Watanabe,2021_PRL_Saryal}. Studying this interplay further and designing optimal cycles achieving different objectives is of interest for future work. Similarly, we focused on the problem of finding optimal, unconstrained cycles in thermodynamic space, such that control parameters, the temperature of the heat bath and mechanical controls, are allowed to vary continuously in time. Although this regime has been studied extensively \cite{2015_PRX_Brandner,2020_PRL_Brandner,2021_PRE_Miangolarra_1,2022_PRE_Frim,2022_PRR_Watanabe} and is experimentally accessible \cite{2016_NatPhys_Martinez}, it is a distinct and worthwhile question to address the problem of constructing optimal, finite-time engines under other more constrained control settings where such smooth variations may not be possible.

In addition and more generally, %
isoperimetric inequalities 
have been a significant direction of study  in mathematics,
and we expect 
them to have important implications in thermodynamic geometry.
We are encouraged by their recent application to the Brownian gyrator~\cite{2021_PRE_Miangolarra_2} and adiabatic thermal engines operating between two heat baths~\cite{2022_PRXQ_alonso}, but we anticipate there remains a great deal to be learned by their application in various thermodynamic settings. In particular, whereas ultimately our bound relied on the introduction of thermodynamic area, isoperimetric inequalities for manifolds with density \cite{morgan_2,rosales,morgan,SM} could lead to further strict bounds on dissipation directly given a work output.

Finally, %
although our main focus here was on classical thermal systems, thermodynamic geometry is equally applicable to quantum settings and this bound likewise should constrain the efficiency of quantum heat engines, a major focus of current
research~\cite{2014_ARPC_Kosloff,2015_PRE_Gardas,2018_entropy_deffner,2020_abiuso_entropy,2020_abiuso_prl,2021_miller_prl}.

\indent \textit{Conclusion}. Here, we have used classical geometric results in concert with geometric approaches to thermodynamics to place a bound on the efficiency of any irreversible heat engine and study it in the specific case of the parametric harmonic oscillator. This bound applies irrespective of the system details or dynamics 
and it suggests new design principles for construction of efficient engines at microscopic scales. %

\begin{acknowledgments}
\textit{Acknowledgments.} %
The authors thank Jamie Simon, Adrianne Zhong, Mart\'{i} Perarnau-Llobet, and Aaron Slipper for many useful discussions and thank Gentaro Watanabe and Yuki Minami for comments on the manuscript. The authors 
also acknowledge the attendees of the 2021 Telluride Information Engines Workshop for useful comments at preliminary stages of this work. AGF is supported by the NSF GRFP under Grant No. DGE 1752814. %
This work was supported in part by the U.S. Army Research Laboratory and the U.S. Army Research Office under contract W911NF-20-1-0151.

\end{acknowledgments}


\begin{thebibliography}{67}
\expandafter\ifx\csname natexlab\endcsname\relax\def\natexlab#1{#1}\fi
\expandafter\ifx\csname bibnamefont\endcsname\relax
  \def\bibnamefont#1{#1}\fi
\expandafter\ifx\csname bibfnamefont\endcsname\relax
  \def\bibfnamefont#1{#1}\fi
\expandafter\ifx\csname citenamefont\endcsname\relax
  \def\citenamefont#1{#1}\fi
\expandafter\ifx\csname url\endcsname\relax
  \def\url#1{\texttt{#1}}\fi
\expandafter\ifx\csname urlprefix\endcsname\relax\def\urlprefix{URL }\fi
\providecommand{\bibinfo}[2]{#2}
\providecommand{\eprint}[2][]{\url{#2}}

\bibitem[{\citenamefont{Jarzynski}(1997)}]{1997_PRL_Jarzynski}
\bibinfo{author}{\bibfnamefont{C.}~\bibnamefont{Jarzynski}},
  \bibinfo{journal}{Phys. Rev. Lett.} \textbf{\bibinfo{volume}{78}},
  \bibinfo{pages}{2690} (\bibinfo{year}{1997}),
  \urlprefix\url{https://link.aps.org/doi/10.1103/PhysRevLett.78.2690}.

\bibitem[{\citenamefont{Sekimoto}(1998)}]{Sekimoto}
\bibinfo{author}{\bibfnamefont{K.}~\bibnamefont{Sekimoto}},
  \bibinfo{journal}{Progress of Theoretical Physics Supplement}
  \textbf{\bibinfo{volume}{130}}, \bibinfo{pages}{17} (\bibinfo{year}{1998}),
  ISSN \bibinfo{issn}{0375-9687}.

\bibitem[{\citenamefont{Crooks}(1999)}]{1999_PRE_Crooks}
\bibinfo{author}{\bibfnamefont{G.~E.} \bibnamefont{Crooks}},
  \bibinfo{journal}{Phys. Rev. E} \textbf{\bibinfo{volume}{60}},
  \bibinfo{pages}{2721} (\bibinfo{year}{1999}),
  \urlprefix\url{https://link.aps.org/doi/10.1103/PhysRevE.60.2721}.

\bibitem[{\citenamefont{Seifert}(2005)}]{2005_PRL_Seifert}
\bibinfo{author}{\bibfnamefont{U.}~\bibnamefont{Seifert}},
  \bibinfo{journal}{Phys. Rev. Lett.} \textbf{\bibinfo{volume}{95}},
  \bibinfo{pages}{040602} (\bibinfo{year}{2005}),
  \urlprefix\url{https://link.aps.org/doi/10.1103/PhysRevLett.95.040602}.

\bibitem[{\citenamefont{Esposito and Van~den
  Broeck}(2010)}]{2010_PRL_Esposito_three}
\bibinfo{author}{\bibfnamefont{M.}~\bibnamefont{Esposito}} \bibnamefont{and}
  \bibinfo{author}{\bibfnamefont{C.}~\bibnamefont{Van~den Broeck}},
  \bibinfo{journal}{Phys. Rev. Lett.} \textbf{\bibinfo{volume}{104}},
  \bibinfo{pages}{090601} (\bibinfo{year}{2010}),
  \urlprefix\url{https://link.aps.org/doi/10.1103/PhysRevLett.104.090601}.

\bibitem[{\citenamefont{Sagawa and Ueda}(2010)}]{2010_PRL_Sagawa}
\bibinfo{author}{\bibfnamefont{T.}~\bibnamefont{Sagawa}} \bibnamefont{and}
  \bibinfo{author}{\bibfnamefont{M.}~\bibnamefont{Ueda}},
  \bibinfo{journal}{Phys. Rev. Lett.} \textbf{\bibinfo{volume}{104}},
  \bibinfo{pages}{090602} (\bibinfo{year}{2010}),
  \urlprefix\url{https://link.aps.org/doi/10.1103/PhysRevLett.104.090602}.

\bibitem[{\citenamefont{Seifert}(2012)}]{2012_RPP_Seifert}
\bibinfo{author}{\bibfnamefont{U.}~\bibnamefont{Seifert}},
  \bibinfo{journal}{Reports on Progress in Physics}
  \textbf{\bibinfo{volume}{75}}, \bibinfo{pages}{126001}
  (\bibinfo{year}{2012}),
  \urlprefix\url{https://doi.org/10.1088/0034-4885/75/12/126001}.

\bibitem[{\citenamefont{Curzon and Ahlborn}(1975)}]{1975_AJP_Curzon}
\bibinfo{author}{\bibfnamefont{F.~L.} \bibnamefont{Curzon}} \bibnamefont{and}
  \bibinfo{author}{\bibfnamefont{B.}~\bibnamefont{Ahlborn}},
  \bibinfo{journal}{American Journal of Physics} \textbf{\bibinfo{volume}{43}},
  \bibinfo{pages}{22} (\bibinfo{year}{1975}).

\bibitem[{\citenamefont{Van~den Broeck}(2005)}]{2005_PRL_VandenBroeck}
\bibinfo{author}{\bibfnamefont{C.}~\bibnamefont{Van~den Broeck}},
  \bibinfo{journal}{Phys. Rev. Lett.} \textbf{\bibinfo{volume}{95}},
  \bibinfo{pages}{190602} (\bibinfo{year}{2005}),
  \urlprefix\url{https://link.aps.org/doi/10.1103/PhysRevLett.95.190602}.

\bibitem[{\citenamefont{Schmiedl and
  Seifert}(2007{\natexlab{a}})}]{2007_EPL_Schmiedl}
\bibinfo{author}{\bibfnamefont{T.}~\bibnamefont{Schmiedl}} \bibnamefont{and}
  \bibinfo{author}{\bibfnamefont{U.}~\bibnamefont{Seifert}},
  \bibinfo{journal}{{EPL} (Europhysics Letters)} \textbf{\bibinfo{volume}{81}},
  \bibinfo{pages}{20003} (\bibinfo{year}{2007}{\natexlab{a}}),
  \urlprefix\url{https://doi.org/10.1209/0295-5075/81/20003}.

\bibitem[{\citenamefont{Esposito et~al.}(2009)\citenamefont{Esposito,
  Lindenberg, and Van~den Broeck}}]{2009_PRL_Esposito}
\bibinfo{author}{\bibfnamefont{M.}~\bibnamefont{Esposito}},
  \bibinfo{author}{\bibfnamefont{K.}~\bibnamefont{Lindenberg}},
  \bibnamefont{and} \bibinfo{author}{\bibfnamefont{C.}~\bibnamefont{Van~den
  Broeck}}, \bibinfo{journal}{Phys. Rev. Lett.} \textbf{\bibinfo{volume}{102}},
  \bibinfo{pages}{130602} (\bibinfo{year}{2009}),
  \urlprefix\url{https://link.aps.org/doi/10.1103/PhysRevLett.102.130602}.

\bibitem[{\citenamefont{Esposito et~al.}(2010)\citenamefont{Esposito, Kawai,
  Lindenberg, and Van~den Broeck}}]{2010_PRL_Esposito}
\bibinfo{author}{\bibfnamefont{M.}~\bibnamefont{Esposito}},
  \bibinfo{author}{\bibfnamefont{R.}~\bibnamefont{Kawai}},
  \bibinfo{author}{\bibfnamefont{K.}~\bibnamefont{Lindenberg}},
  \bibnamefont{and} \bibinfo{author}{\bibfnamefont{C.}~\bibnamefont{Van~den
  Broeck}}, \bibinfo{journal}{Phys. Rev. Lett.} \textbf{\bibinfo{volume}{105}},
  \bibinfo{pages}{150603} (\bibinfo{year}{2010}),
  \urlprefix\url{https://link.aps.org/doi/10.1103/PhysRevLett.105.150603}.

\bibitem[{\citenamefont{Brandner et~al.}(2015)\citenamefont{Brandner, Saito,
  and Seifert}}]{2015_PRX_Brandner}
\bibinfo{author}{\bibfnamefont{K.}~\bibnamefont{Brandner}},
  \bibinfo{author}{\bibfnamefont{K.}~\bibnamefont{Saito}}, \bibnamefont{and}
  \bibinfo{author}{\bibfnamefont{U.}~\bibnamefont{Seifert}},
  \bibinfo{journal}{Phys. Rev. X} \textbf{\bibinfo{volume}{5}},
  \bibinfo{pages}{031019} (\bibinfo{year}{2015}),
  \urlprefix\url{https://link.aps.org/doi/10.1103/PhysRevX.5.031019}.

\bibitem[{\citenamefont{Proesmans et~al.}(2016)\citenamefont{Proesmans,
  Cleuren, and Van~den Broeck}}]{2016_PRL_Proesmans}
\bibinfo{author}{\bibfnamefont{K.}~\bibnamefont{Proesmans}},
  \bibinfo{author}{\bibfnamefont{B.}~\bibnamefont{Cleuren}}, \bibnamefont{and}
  \bibinfo{author}{\bibfnamefont{C.}~\bibnamefont{Van~den Broeck}},
  \bibinfo{journal}{Phys. Rev. Lett.} \textbf{\bibinfo{volume}{116}},
  \bibinfo{pages}{220601} (\bibinfo{year}{2016}),
  \urlprefix\url{https://link.aps.org/doi/10.1103/PhysRevLett.116.220601}.

\bibitem[{\citenamefont{Shiraishi et~al.}(2016)\citenamefont{Shiraishi, Saito,
  and Tasaki}}]{2016_PRL_Shiraishi}
\bibinfo{author}{\bibfnamefont{N.}~\bibnamefont{Shiraishi}},
  \bibinfo{author}{\bibfnamefont{K.}~\bibnamefont{Saito}}, \bibnamefont{and}
  \bibinfo{author}{\bibfnamefont{H.}~\bibnamefont{Tasaki}},
  \bibinfo{journal}{Phys. Rev. Lett.} \textbf{\bibinfo{volume}{117}},
  \bibinfo{pages}{190601} (\bibinfo{year}{2016}),
  \urlprefix\url{https://link.aps.org/doi/10.1103/PhysRevLett.117.190601}.

\bibitem[{\citenamefont{Ma et~al.}(2018{\natexlab{a}})\citenamefont{Ma, Xu,
  Dong, and Sun}}]{2018_PRE_Ma_1}
\bibinfo{author}{\bibfnamefont{Y.-H.} \bibnamefont{Ma}},
  \bibinfo{author}{\bibfnamefont{D.}~\bibnamefont{Xu}},
  \bibinfo{author}{\bibfnamefont{H.}~\bibnamefont{Dong}}, \bibnamefont{and}
  \bibinfo{author}{\bibfnamefont{C.-P.} \bibnamefont{Sun}},
  \bibinfo{journal}{Phys. Rev. E} \textbf{\bibinfo{volume}{98}},
  \bibinfo{pages}{022133} (\bibinfo{year}{2018}{\natexlab{a}}),
  \urlprefix\url{https://link.aps.org/doi/10.1103/PhysRevE.98.022133}.

\bibitem[{\citenamefont{Ma et~al.}(2018{\natexlab{b}})\citenamefont{Ma, Xu,
  Dong, and Sun}}]{2018_PRE_Ma_2}
\bibinfo{author}{\bibfnamefont{Y.-H.} \bibnamefont{Ma}},
  \bibinfo{author}{\bibfnamefont{D.}~\bibnamefont{Xu}},
  \bibinfo{author}{\bibfnamefont{H.}~\bibnamefont{Dong}}, \bibnamefont{and}
  \bibinfo{author}{\bibfnamefont{C.-P.} \bibnamefont{Sun}},
  \bibinfo{journal}{Phys. Rev. E} \textbf{\bibinfo{volume}{98}},
  \bibinfo{pages}{042112} (\bibinfo{year}{2018}{\natexlab{b}}),
  \urlprefix\url{https://link.aps.org/doi/10.1103/PhysRevE.98.042112}.

\bibitem[{\citenamefont{Abiuso and Perarnau-Llobet}(2020)}]{2020_abiuso_prl}
\bibinfo{author}{\bibfnamefont{P.}~\bibnamefont{Abiuso}} \bibnamefont{and}
  \bibinfo{author}{\bibfnamefont{M.}~\bibnamefont{Perarnau-Llobet}},
  \bibinfo{journal}{Phys. Rev. Lett.} \textbf{\bibinfo{volume}{124}},
  \bibinfo{pages}{110606} (\bibinfo{year}{2020}),
  \urlprefix\url{https://link.aps.org/doi/10.1103/PhysRevLett.124.110606}.

\bibitem[{\citenamefont{Ma et~al.}(2020)\citenamefont{Ma, Zhai, Chen, Sun, and
  Dong}}]{2020_PRL_Ma}
\bibinfo{author}{\bibfnamefont{Y.-H.} \bibnamefont{Ma}},
  \bibinfo{author}{\bibfnamefont{R.-X.} \bibnamefont{Zhai}},
  \bibinfo{author}{\bibfnamefont{J.}~\bibnamefont{Chen}},
  \bibinfo{author}{\bibfnamefont{C.~P.} \bibnamefont{Sun}}, \bibnamefont{and}
  \bibinfo{author}{\bibfnamefont{H.}~\bibnamefont{Dong}},
  \bibinfo{journal}{Phys. Rev. Lett.} \textbf{\bibinfo{volume}{125}},
  \bibinfo{pages}{210601} (\bibinfo{year}{2020}),
  \urlprefix\url{https://link.aps.org/doi/10.1103/PhysRevLett.125.210601}.

\bibitem[{\citenamefont{Miller and Mehboudi}(2020)}]{2020_PRL_Miller}
\bibinfo{author}{\bibfnamefont{H.~J.~D.} \bibnamefont{Miller}}
  \bibnamefont{and} \bibinfo{author}{\bibfnamefont{M.}~\bibnamefont{Mehboudi}},
  \bibinfo{journal}{Phys. Rev. Lett.} \textbf{\bibinfo{volume}{125}},
  \bibinfo{pages}{260602} (\bibinfo{year}{2020}),
  \urlprefix\url{https://link.aps.org/doi/10.1103/PhysRevLett.125.260602}.

\bibitem[{\citenamefont{Movilla~Miangolarra
  et~al.}(2021{\natexlab{a}})\citenamefont{Movilla~Miangolarra, Fu, Taghvaei,
  Chen, and Georgiou}}]{2021_PRE_Miangolarra_1}
\bibinfo{author}{\bibfnamefont{O.}~\bibnamefont{Movilla~Miangolarra}},
  \bibinfo{author}{\bibfnamefont{R.}~\bibnamefont{Fu}},
  \bibinfo{author}{\bibfnamefont{A.}~\bibnamefont{Taghvaei}},
  \bibinfo{author}{\bibfnamefont{Y.}~\bibnamefont{Chen}}, \bibnamefont{and}
  \bibinfo{author}{\bibfnamefont{T.~T.} \bibnamefont{Georgiou}},
  \bibinfo{journal}{Phys. Rev. E} \textbf{\bibinfo{volume}{103}},
  \bibinfo{pages}{062103} (\bibinfo{year}{2021}{\natexlab{a}}),
  \urlprefix\url{https://link.aps.org/doi/10.1103/PhysRevE.103.062103}.

\bibitem[{\citenamefont{Frim and DeWeese}(2022)}]{2022_PRE_Frim}
\bibinfo{author}{\bibfnamefont{A.~G.} \bibnamefont{Frim}} \bibnamefont{and}
  \bibinfo{author}{\bibfnamefont{M.~R.} \bibnamefont{DeWeese}},
  \bibinfo{journal}{Phys. Rev. E} \textbf{\bibinfo{volume}{105}},
  \bibinfo{pages}{L052103} (\bibinfo{year}{2022}),
  \urlprefix\url{https://link.aps.org/doi/10.1103/PhysRevE.105.L052103}.

\bibitem[{\citenamefont{Watanabe and Minami}(2022)}]{2022_PRR_Watanabe}
\bibinfo{author}{\bibfnamefont{G.}~\bibnamefont{Watanabe}} \bibnamefont{and}
  \bibinfo{author}{\bibfnamefont{Y.}~\bibnamefont{Minami}},
  \bibinfo{journal}{Phys. Rev. Research} \textbf{\bibinfo{volume}{4}},
  \bibinfo{pages}{L012008} (\bibinfo{year}{2022}),
  \urlprefix\url{https://link.aps.org/doi/10.1103/PhysRevResearch.4.L012008}.

\bibitem[{\citenamefont{Terr\'en~Alonso
  et~al.}(2022)\citenamefont{Terr\'en~Alonso, Abiuso, Perarnau-Llobet, and
  Arrachea}}]{2022_PRXQ_alonso}
\bibinfo{author}{\bibfnamefont{P.}~\bibnamefont{Terr\'en~Alonso}},
  \bibinfo{author}{\bibfnamefont{P.}~\bibnamefont{Abiuso}},
  \bibinfo{author}{\bibfnamefont{M.}~\bibnamefont{Perarnau-Llobet}},
  \bibnamefont{and} \bibinfo{author}{\bibfnamefont{L.}~\bibnamefont{Arrachea}},
  \bibinfo{journal}{PRX Quantum} \textbf{\bibinfo{volume}{3}},
  \bibinfo{pages}{010326} (\bibinfo{year}{2022}),
  \urlprefix\url{https://link.aps.org/doi/10.1103/PRXQuantum.3.010326}.

\bibitem[{\citenamefont{Mart{\'\i}nez et~al.}(2016)\citenamefont{Mart{\'\i}nez,
  Rold{\'a}n, Dinis, Petrov, Parrondo, and Rica}}]{2016_NatPhys_Martinez}
\bibinfo{author}{\bibfnamefont{I.~A.} \bibnamefont{Mart{\'\i}nez}},
  \bibinfo{author}{\bibfnamefont{{\'E}.}~\bibnamefont{Rold{\'a}n}},
  \bibinfo{author}{\bibfnamefont{L.}~\bibnamefont{Dinis}},
  \bibinfo{author}{\bibfnamefont{D.}~\bibnamefont{Petrov}},
  \bibinfo{author}{\bibfnamefont{J.~M.~R.} \bibnamefont{Parrondo}},
  \bibnamefont{and} \bibinfo{author}{\bibfnamefont{R.~A.} \bibnamefont{Rica}},
  \bibinfo{journal}{Nature Physics} \textbf{\bibinfo{volume}{12}},
  \bibinfo{pages}{67} (\bibinfo{year}{2016}),
  \urlprefix\url{https://doi.org/10.1038/nphys3518}.

\bibitem[{\citenamefont{Brandner and Saito}(2020)}]{2020_PRL_Brandner}
\bibinfo{author}{\bibfnamefont{K.}~\bibnamefont{Brandner}} \bibnamefont{and}
  \bibinfo{author}{\bibfnamefont{K.}~\bibnamefont{Saito}},
  \bibinfo{journal}{Phys. Rev. Lett.} \textbf{\bibinfo{volume}{124}},
  \bibinfo{pages}{040602} (\bibinfo{year}{2020}),
  \urlprefix\url{https://link.aps.org/doi/10.1103/PhysRevLett.124.040602}.

\bibitem[{\citenamefont{Movilla~Miangolarra
  et~al.}(2021{\natexlab{b}})\citenamefont{Movilla~Miangolarra, Taghvaei, Fu,
  Chen, and Georgiou}}]{2021_PRE_Miangolarra_2}
\bibinfo{author}{\bibfnamefont{O.}~\bibnamefont{Movilla~Miangolarra}},
  \bibinfo{author}{\bibfnamefont{A.}~\bibnamefont{Taghvaei}},
  \bibinfo{author}{\bibfnamefont{R.}~\bibnamefont{Fu}},
  \bibinfo{author}{\bibfnamefont{Y.}~\bibnamefont{Chen}}, \bibnamefont{and}
  \bibinfo{author}{\bibfnamefont{T.~T.} \bibnamefont{Georgiou}},
  \bibinfo{journal}{Phys. Rev. E} \textbf{\bibinfo{volume}{104}},
  \bibinfo{pages}{044101} (\bibinfo{year}{2021}{\natexlab{b}}),
  \urlprefix\url{https://link.aps.org/doi/10.1103/PhysRevE.104.044101}.

\bibitem[{\citenamefont{Pietzonka et~al.}(2016)\citenamefont{Pietzonka, Barato,
  and Seifert}}]{2016_JSM_Pietzonka}
\bibinfo{author}{\bibfnamefont{P.}~\bibnamefont{Pietzonka}},
  \bibinfo{author}{\bibfnamefont{A.~C.} \bibnamefont{Barato}},
  \bibnamefont{and} \bibinfo{author}{\bibfnamefont{U.}~\bibnamefont{Seifert}},
  \bibinfo{journal}{Journal of Statistical Mechanics: Theory and Experiment}
  \textbf{\bibinfo{volume}{2016}}, \bibinfo{pages}{124004}
  (\bibinfo{year}{2016}),
  \urlprefix\url{https://doi.org/10.1088/1742-5468/2016/12/124004}.

\bibitem[{\citenamefont{Pietzonka et~al.}(2019)\citenamefont{Pietzonka, Fodor,
  Lohrmann, Cates, and Seifert}}]{2019_PRX_Pietzonka}
\bibinfo{author}{\bibfnamefont{P.}~\bibnamefont{Pietzonka}},
  \bibinfo{author}{\bibfnamefont{E.}~\bibnamefont{Fodor}},
  \bibinfo{author}{\bibfnamefont{C.}~\bibnamefont{Lohrmann}},
  \bibinfo{author}{\bibfnamefont{M.~E.} \bibnamefont{Cates}}, \bibnamefont{and}
  \bibinfo{author}{\bibfnamefont{U.}~\bibnamefont{Seifert}},
  \bibinfo{journal}{Phys. Rev. X} \textbf{\bibinfo{volume}{9}},
  \bibinfo{pages}{041032} (\bibinfo{year}{2019}),
  \urlprefix\url{https://link.aps.org/doi/10.1103/PhysRevX.9.041032}.

\bibitem[{\citenamefont{Ekeh et~al.}(2020)\citenamefont{Ekeh, Cates, and
  Fodor}}]{2020_PRE_Ekeh}
\bibinfo{author}{\bibfnamefont{T.}~\bibnamefont{Ekeh}},
  \bibinfo{author}{\bibfnamefont{M.~E.} \bibnamefont{Cates}}, \bibnamefont{and}
  \bibinfo{author}{\bibfnamefont{E.}~\bibnamefont{Fodor}},
  \bibinfo{journal}{Phys. Rev. E} \textbf{\bibinfo{volume}{102}},
  \bibinfo{pages}{010101} (\bibinfo{year}{2020}),
  \urlprefix\url{https://link.aps.org/doi/10.1103/PhysRevE.102.010101}.

\bibitem[{\citenamefont{Lee et~al.}(2020)\citenamefont{Lee, Park, and
  Park}}]{2020_PRE_Lee}
\bibinfo{author}{\bibfnamefont{J.~S.} \bibnamefont{Lee}},
  \bibinfo{author}{\bibfnamefont{J.-M.} \bibnamefont{Park}}, \bibnamefont{and}
  \bibinfo{author}{\bibfnamefont{H.}~\bibnamefont{Park}},
  \bibinfo{journal}{Phys. Rev. E} \textbf{\bibinfo{volume}{102}},
  \bibinfo{pages}{032116} (\bibinfo{year}{2020}),
  \urlprefix\url{https://link.aps.org/doi/10.1103/PhysRevE.102.032116}.

\bibitem[{\citenamefont{Weinhold}(1975)}]{1975_JCP_Weinhold}
\bibinfo{author}{\bibfnamefont{F.}~\bibnamefont{Weinhold}},
  \bibinfo{journal}{The Journal of Chemical Physics}
  \textbf{\bibinfo{volume}{63}}, \bibinfo{pages}{2479} (\bibinfo{year}{1975}),
  \eprint{https://doi.org/10.1063/1.431689},
  \urlprefix\url{https://doi.org/10.1063/1.431689}.

\bibitem[{\citenamefont{Ruppeiner}(1979)}]{1979_PRA_Ruppeiner}
\bibinfo{author}{\bibfnamefont{G.}~\bibnamefont{Ruppeiner}},
  \bibinfo{journal}{Phys. Rev. A} \textbf{\bibinfo{volume}{20}},
  \bibinfo{pages}{1608} (\bibinfo{year}{1979}),
  \urlprefix\url{https://link.aps.org/doi/10.1103/PhysRevA.20.1608}.

\bibitem[{\citenamefont{Salamon and Berry}(1983)}]{1983_Salamon_PRL}
\bibinfo{author}{\bibfnamefont{P.}~\bibnamefont{Salamon}} \bibnamefont{and}
  \bibinfo{author}{\bibfnamefont{R.~S.} \bibnamefont{Berry}},
  \bibinfo{journal}{Phys. Rev. Lett.} \textbf{\bibinfo{volume}{51}},
  \bibinfo{pages}{1127} (\bibinfo{year}{1983}),
  \urlprefix\url{https://link.aps.org/doi/10.1103/PhysRevLett.51.1127}.

\bibitem[{\citenamefont{Salamon et~al.}(1984)\citenamefont{Salamon, Nulton, and
  Ihrig}}]{1984_JCP_Salamon}
\bibinfo{author}{\bibfnamefont{P.}~\bibnamefont{Salamon}},
  \bibinfo{author}{\bibfnamefont{J.}~\bibnamefont{Nulton}}, \bibnamefont{and}
  \bibinfo{author}{\bibfnamefont{E.}~\bibnamefont{Ihrig}},
  \bibinfo{journal}{The Journal of Chemical Physics}
  \textbf{\bibinfo{volume}{80}}, \bibinfo{pages}{436} (\bibinfo{year}{1984}),
  \eprint{https://doi.org/10.1063/1.446467},
  \urlprefix\url{https://doi.org/10.1063/1.446467}.

\bibitem[{\citenamefont{Schl{\"o}gl}(1985)}]{1985_ZPB_Schlogl}
\bibinfo{author}{\bibfnamefont{F.}~\bibnamefont{Schl{\"o}gl}},
  \bibinfo{journal}{Zeitschrift f{\"u}r Physik B Condensed Matter}
  \textbf{\bibinfo{volume}{59}}, \bibinfo{pages}{449} (\bibinfo{year}{1985}),
  \urlprefix\url{https://doi.org/10.1007/BF01328857}.

\bibitem[{\citenamefont{Ruppeiner}(1995)}]{1995_RMP_Ruppeiner}
\bibinfo{author}{\bibfnamefont{G.}~\bibnamefont{Ruppeiner}},
  \bibinfo{journal}{Rev. Mod. Phys.} \textbf{\bibinfo{volume}{67}},
  \bibinfo{pages}{605} (\bibinfo{year}{1995}),
  \urlprefix\url{https://link.aps.org/doi/10.1103/RevModPhys.67.605}.

\bibitem[{\citenamefont{Crooks}(2007)}]{2007_PRL_Crooks}
\bibinfo{author}{\bibfnamefont{G.~E.} \bibnamefont{Crooks}},
  \bibinfo{journal}{Phys. Rev. Lett.} \textbf{\bibinfo{volume}{99}},
  \bibinfo{pages}{100602} (\bibinfo{year}{2007}),
  \urlprefix\url{https://link.aps.org/doi/10.1103/PhysRevLett.99.100602}.

\bibitem[{\citenamefont{Sivak and Crooks}(2012)}]{2012_PRL_Sivak}
\bibinfo{author}{\bibfnamefont{D.~A.} \bibnamefont{Sivak}} \bibnamefont{and}
  \bibinfo{author}{\bibfnamefont{G.~E.} \bibnamefont{Crooks}},
  \bibinfo{journal}{Phys. Rev. Lett.} \textbf{\bibinfo{volume}{108}},
  \bibinfo{pages}{190602} (\bibinfo{year}{2012}),
  \urlprefix\url{https://link.aps.org/doi/10.1103/PhysRevLett.108.190602}.

\bibitem[{\citenamefont{Zulkowski et~al.}(2012)\citenamefont{Zulkowski, Sivak,
  Crooks, and DeWeese}}]{2012_PRE_zulkowski}
\bibinfo{author}{\bibfnamefont{P.~R.} \bibnamefont{Zulkowski}},
  \bibinfo{author}{\bibfnamefont{D.~A.} \bibnamefont{Sivak}},
  \bibinfo{author}{\bibfnamefont{G.~E.} \bibnamefont{Crooks}},
  \bibnamefont{and} \bibinfo{author}{\bibfnamefont{M.~R.}
  \bibnamefont{DeWeese}}, \bibinfo{journal}{Phys. Rev. E}
  \textbf{\bibinfo{volume}{86}}, \bibinfo{pages}{041148}
  (\bibinfo{year}{2012}),
  \urlprefix\url{https://link.aps.org/doi/10.1103/PhysRevE.86.041148}.

\bibitem[{\citenamefont{Zulkowski and
  DeWeese}(2014)}]{2014_PRE_Zulkowski_erasure}
\bibinfo{author}{\bibfnamefont{P.~R.} \bibnamefont{Zulkowski}}
  \bibnamefont{and} \bibinfo{author}{\bibfnamefont{M.~R.}
  \bibnamefont{DeWeese}}, \bibinfo{journal}{Phys. Rev. E}
  \textbf{\bibinfo{volume}{89}}, \bibinfo{pages}{052140}
  (\bibinfo{year}{2014}),
  \urlprefix\url{https://link.aps.org/doi/10.1103/PhysRevE.89.052140}.

\bibitem[{\citenamefont{Zulkowski et~al.}(2013)\citenamefont{Zulkowski, Sivak,
  and DeWeese}}]{2013_PLOS_Zulkowski}
\bibinfo{author}{\bibfnamefont{P.~R.} \bibnamefont{Zulkowski}},
  \bibinfo{author}{\bibfnamefont{D.~A.} \bibnamefont{Sivak}}, \bibnamefont{and}
  \bibinfo{author}{\bibfnamefont{M.~R.} \bibnamefont{DeWeese}},
  \bibinfo{journal}{PLOS ONE} \textbf{\bibinfo{volume}{8}}, \bibinfo{pages}{1}
  (\bibinfo{year}{2013}),
  \urlprefix\url{https://doi.org/10.1371/journal.pone.0082754}.

\bibitem[{\citenamefont{Zulkowski and
  DeWeese}(2015{\natexlab{a}})}]{2015_PRE_Zulkowski_quantum}
\bibinfo{author}{\bibfnamefont{P.~R.} \bibnamefont{Zulkowski}}
  \bibnamefont{and} \bibinfo{author}{\bibfnamefont{M.~R.}
  \bibnamefont{DeWeese}}, \bibinfo{journal}{Phys. Rev. E}
  \textbf{\bibinfo{volume}{92}}, \bibinfo{pages}{032113}
  (\bibinfo{year}{2015}{\natexlab{a}}),
  \urlprefix\url{https://link.aps.org/doi/10.1103/PhysRevE.92.032113}.

\bibitem[{\citenamefont{Zulkowski and
  DeWeese}(2015{\natexlab{b}})}]{2015_PRE_zulkowski_over}
\bibinfo{author}{\bibfnamefont{P.~R.} \bibnamefont{Zulkowski}}
  \bibnamefont{and} \bibinfo{author}{\bibfnamefont{M.~R.}
  \bibnamefont{DeWeese}}, \bibinfo{journal}{Physical Review E}
  \textbf{\bibinfo{volume}{92}}, \bibinfo{pages}{032117}
  (\bibinfo{year}{2015}{\natexlab{b}}), ISSN \bibinfo{issn}{1539-3755,
  1550-2376},
  \urlprefix\url{https://link.aps.org/doi/10.1103/PhysRevE.92.032117}.

\bibitem[{\citenamefont{Rotskoff and Crooks}(2015)}]{2015_PRE_Rotskoff}
\bibinfo{author}{\bibfnamefont{G.~M.} \bibnamefont{Rotskoff}} \bibnamefont{and}
  \bibinfo{author}{\bibfnamefont{G.~E.} \bibnamefont{Crooks}},
  \bibinfo{journal}{Phys. Rev. E} \textbf{\bibinfo{volume}{92}},
  \bibinfo{pages}{060102} (\bibinfo{year}{2015}),
  \urlprefix\url{https://link.aps.org/doi/10.1103/PhysRevE.92.060102}.

\bibitem[{\citenamefont{Abiuso et~al.}(2020)\citenamefont{Abiuso, Miller,
  Perarnau-Llobet, and Scandi}}]{2020_abiuso_entropy}
\bibinfo{author}{\bibfnamefont{P.}~\bibnamefont{Abiuso}},
  \bibinfo{author}{\bibfnamefont{H.~J.~D.} \bibnamefont{Miller}},
  \bibinfo{author}{\bibfnamefont{M.}~\bibnamefont{Perarnau-Llobet}},
  \bibnamefont{and} \bibinfo{author}{\bibfnamefont{M.}~\bibnamefont{Scandi}},
  \bibinfo{journal}{Entropy} \textbf{\bibinfo{volume}{22}}
  (\bibinfo{year}{2020}), ISSN \bibinfo{issn}{1099-4300},
  \urlprefix\url{https://www.mdpi.com/1099-4300/22/10/1076}.

\bibitem[{\citenamefont{Blaber and Sivak}(2020)}]{2020_JCP_Blaber}
\bibinfo{author}{\bibfnamefont{S.}~\bibnamefont{Blaber}} \bibnamefont{and}
  \bibinfo{author}{\bibfnamefont{D.~A.} \bibnamefont{Sivak}},
  \bibinfo{journal}{The Journal of Chemical Physics}
  \textbf{\bibinfo{volume}{153}}, \bibinfo{pages}{244119}
  (\bibinfo{year}{2020}).

\bibitem[{\citenamefont{Payne}(1967)}]{payne_isoperimetric_1967}
\bibinfo{author}{\bibfnamefont{L.~E.} \bibnamefont{Payne}},
  \bibinfo{journal}{SIAM Review} \textbf{\bibinfo{volume}{9}},
  \bibinfo{pages}{453} (\bibinfo{year}{1967}), ISSN \bibinfo{issn}{00361445},
  \bibinfo{note}{publisher: Society for Industrial and Applied Mathematics}.

\bibitem[{\citenamefont{Osserman}(1978)}]{Osserman_1978}
\bibinfo{author}{\bibfnamefont{R.}~\bibnamefont{Osserman}},
  \bibinfo{journal}{Bulletin of the American Mathematical Society}
  \textbf{\bibinfo{volume}{84}}, \bibinfo{pages}{1182 } (\bibinfo{year}{1978}),
  \urlprefix\url{https://doi.org/}.

\bibitem[{SM()}]{SM}
\bibinfo{note}{See Supplemental Material at [URL will be inserted by publisher]
  for further details on derivations, numerical methods, isoperimetric bounds
  for arbitrary curvatures, and the potential for manifolds with density}.

\bibitem[{\citenamefont{Schmiedl and
  Seifert}(2007{\natexlab{b}})}]{2007_PRL_Schmiedl}
\bibinfo{author}{\bibfnamefont{T.}~\bibnamefont{Schmiedl}} \bibnamefont{and}
  \bibinfo{author}{\bibfnamefont{U.}~\bibnamefont{Seifert}},
  \bibinfo{journal}{Phys. Rev. Lett.} \textbf{\bibinfo{volume}{98}},
  \bibinfo{pages}{108301} (\bibinfo{year}{2007}{\natexlab{b}}),
  \urlprefix\url{https://link.aps.org/doi/10.1103/PhysRevLett.98.108301}.

\bibitem[{\citenamefont{Blickle and Bechinger}(2012)}]{2011_NatPhys_Blickle}
\bibinfo{author}{\bibfnamefont{V.}~\bibnamefont{Blickle}} \bibnamefont{and}
  \bibinfo{author}{\bibfnamefont{C.}~\bibnamefont{Bechinger}},
  \bibinfo{journal}{Nature Physics} \textbf{\bibinfo{volume}{8}},
  \bibinfo{pages}{143} (\bibinfo{year}{2012}),
  \urlprefix\url{https://doi.org/10.1038/nphys2163}.

\bibitem[{\citenamefont{Mart\'{i}nez et~al.}(2015)\citenamefont{Mart\'{i}nez,
  Rold\'{a}n, Dinis, Petrov, and Rica}}]{2015_PRL_Martinez}
\bibinfo{author}{\bibfnamefont{I.~A.} \bibnamefont{Mart\'{i}nez}},
  \bibinfo{author}{\bibfnamefont{E.}~\bibnamefont{Rold\'{a}n}},
  \bibinfo{author}{\bibfnamefont{L.}~\bibnamefont{Dinis}},
  \bibinfo{author}{\bibfnamefont{D.}~\bibnamefont{Petrov}}, \bibnamefont{and}
  \bibinfo{author}{\bibfnamefont{R.~A.} \bibnamefont{Rica}},
  \bibinfo{journal}{Phys. Rev. Lett.} \textbf{\bibinfo{volume}{114}},
  \bibinfo{pages}{120601} (\bibinfo{year}{2015}),
  \urlprefix\url{https://link.aps.org/doi/10.1103/PhysRevLett.114.120601}.

\bibitem[{\citenamefont{Mart\'{i}nez et~al.}(2017)\citenamefont{Mart\'{i}nez,
  Rold\'{a}n, Dinis, and Rica}}]{2017_SM_Martinez}
\bibinfo{author}{\bibfnamefont{I.~A.} \bibnamefont{Mart\'{i}nez}},
  \bibinfo{author}{\bibfnamefont{E.}~\bibnamefont{Rold\'{a}n}},
  \bibinfo{author}{\bibfnamefont{L.}~\bibnamefont{Dinis}}, \bibnamefont{and}
  \bibinfo{author}{\bibfnamefont{R.~A.} \bibnamefont{Rica}},
  \bibinfo{journal}{Soft Matter} \textbf{\bibinfo{volume}{13}},
  \bibinfo{pages}{22} (\bibinfo{year}{2017}).

\bibitem[{\citenamefont{Huang and Krishnaprasad}(2020)}]{2020_Huang}
\bibinfo{author}{\bibfnamefont{Y.}~\bibnamefont{Huang}} \bibnamefont{and}
  \bibinfo{author}{\bibfnamefont{P.~S.} \bibnamefont{Krishnaprasad}},
  \bibinfo{journal}{Discrete \& Continuous Dynamical Systems - S}
  \textbf{\bibinfo{volume}{13}}, \bibinfo{pages}{1243} (\bibinfo{year}{2020}).

\bibitem[{\citenamefont{Ritor\'{e}}(2001)}]{Ritore}
\bibinfo{author}{\bibfnamefont{M.}~\bibnamefont{Ritor\'{e}}},
  \bibinfo{journal}{Communications in Analysis and Geometry}
  \textbf{\bibinfo{volume}{9}}, \bibinfo{pages}{1093 } (\bibinfo{year}{2001}).

\bibitem[{\citenamefont{Verley et~al.}(2014)\citenamefont{Verley, Willaert,
  Van~den Broeck, and Esposito}}]{2014_PRE_Verley}
\bibinfo{author}{\bibfnamefont{G.}~\bibnamefont{Verley}},
  \bibinfo{author}{\bibfnamefont{T.}~\bibnamefont{Willaert}},
  \bibinfo{author}{\bibfnamefont{C.}~\bibnamefont{Van~den Broeck}},
  \bibnamefont{and} \bibinfo{author}{\bibfnamefont{M.}~\bibnamefont{Esposito}},
  \bibinfo{journal}{Phys. Rev. E} \textbf{\bibinfo{volume}{90}},
  \bibinfo{pages}{052145} (\bibinfo{year}{2014}),
  \urlprefix\url{https://link.aps.org/doi/10.1103/PhysRevE.90.052145}.

\bibitem[{\citenamefont{Ito et~al.}(2019)\citenamefont{Ito, Jiang, and
  Watanabe}}]{2019_arxiv_Ito}
\bibinfo{author}{\bibfnamefont{K.}~\bibnamefont{Ito}},
  \bibinfo{author}{\bibfnamefont{C.}~\bibnamefont{Jiang}}, \bibnamefont{and}
  \bibinfo{author}{\bibfnamefont{G.}~\bibnamefont{Watanabe}}
  (\bibinfo{year}{2019}), \eprint{1910.08096}.

\bibitem[{\citenamefont{Saryal et~al.}(2021)\citenamefont{Saryal, Gerry, Khait,
  Segal, and Agarwalla}}]{2021_PRL_Saryal}
\bibinfo{author}{\bibfnamefont{S.}~\bibnamefont{Saryal}},
  \bibinfo{author}{\bibfnamefont{M.}~\bibnamefont{Gerry}},
  \bibinfo{author}{\bibfnamefont{I.}~\bibnamefont{Khait}},
  \bibinfo{author}{\bibfnamefont{D.}~\bibnamefont{Segal}}, \bibnamefont{and}
  \bibinfo{author}{\bibfnamefont{B.~K.} \bibnamefont{Agarwalla}},
  \bibinfo{journal}{Phys. Rev. Lett.} \textbf{\bibinfo{volume}{127}},
  \bibinfo{pages}{190603} (\bibinfo{year}{2021}),
  \urlprefix\url{https://link.aps.org/doi/10.1103/PhysRevLett.127.190603}.

\bibitem[{\citenamefont{Pietzonka and Seifert}(2018)}]{2018_PRL_Pietzonka}
\bibinfo{author}{\bibfnamefont{P.}~\bibnamefont{Pietzonka}} \bibnamefont{and}
  \bibinfo{author}{\bibfnamefont{U.}~\bibnamefont{Seifert}},
  \bibinfo{journal}{Phys. Rev. Lett.} \textbf{\bibinfo{volume}{120}},
  \bibinfo{pages}{190602} (\bibinfo{year}{2018}),
  \urlprefix\url{https://link.aps.org/doi/10.1103/PhysRevLett.120.190602}.

\bibitem[{\citenamefont{Morgan}(2005)}]{morgan_2}
\bibinfo{author}{\bibfnamefont{F.}~\bibnamefont{Morgan}},
  \bibinfo{journal}{Notices of the American Mathematical Society}
  \textbf{\bibinfo{volume}{52}}, \bibinfo{pages}{853} (\bibinfo{year}{2005}).

\bibitem[{\citenamefont{Rosales et~al.}(2008)\citenamefont{Rosales, Ca{\~n}ete,
  Bayle, and Morgan}}]{rosales}
\bibinfo{author}{\bibfnamefont{C.}~\bibnamefont{Rosales}},
  \bibinfo{author}{\bibfnamefont{A.}~\bibnamefont{Ca{\~n}ete}},
  \bibinfo{author}{\bibfnamefont{V.}~\bibnamefont{Bayle}}, \bibnamefont{and}
  \bibinfo{author}{\bibfnamefont{F.}~\bibnamefont{Morgan}},
  \bibinfo{journal}{Calculus of Variations and Partial Differential Equations}
  \textbf{\bibinfo{volume}{31}}, \bibinfo{pages}{27} (\bibinfo{year}{2008}),
  \urlprefix\url{https://doi.org/10.1007/s00526-007-0104-y}.

\bibitem[{\citenamefont{Morgan and Pratelli}(2013)}]{morgan}
\bibinfo{author}{\bibfnamefont{F.}~\bibnamefont{Morgan}} \bibnamefont{and}
  \bibinfo{author}{\bibfnamefont{A.}~\bibnamefont{Pratelli}},
  \bibinfo{journal}{Annals of Global Analysis and Geometry}
  \textbf{\bibinfo{volume}{43}}, \bibinfo{pages}{331} (\bibinfo{year}{2013}),
  \urlprefix\url{https://doi.org/10.1007/s10455-012-9348-7}.

\bibitem[{\citenamefont{Kosloff and Levy}(2014)}]{2014_ARPC_Kosloff}
\bibinfo{author}{\bibfnamefont{R.}~\bibnamefont{Kosloff}} \bibnamefont{and}
  \bibinfo{author}{\bibfnamefont{A.}~\bibnamefont{Levy}},
  \bibinfo{journal}{Annual Review of Physical Chemistry}
  \textbf{\bibinfo{volume}{65}}, \bibinfo{pages}{365} (\bibinfo{year}{2014}),
  \bibinfo{note}{pMID: 24689798},
  \eprint{https://doi.org/10.1146/annurev-physchem-040513-103724},
  \urlprefix\url{https://doi.org/10.1146/annurev-physchem-040513-103724}.

\bibitem[{\citenamefont{Gardas and Deffner}(2015)}]{2015_PRE_Gardas}
\bibinfo{author}{\bibfnamefont{B.}~\bibnamefont{Gardas}} \bibnamefont{and}
  \bibinfo{author}{\bibfnamefont{S.}~\bibnamefont{Deffner}},
  \bibinfo{journal}{Phys. Rev. E} \textbf{\bibinfo{volume}{92}},
  \bibinfo{pages}{042126} (\bibinfo{year}{2015}),
  \urlprefix\url{https://link.aps.org/doi/10.1103/PhysRevE.92.042126}.

\bibitem[{\citenamefont{Deffner}(2018)}]{2018_entropy_deffner}
\bibinfo{author}{\bibfnamefont{S.}~\bibnamefont{Deffner}},
  \bibinfo{journal}{Entropy} \textbf{\bibinfo{volume}{20}}
  (\bibinfo{year}{2018}), ISSN \bibinfo{issn}{1099-4300},
  \urlprefix\url{https://www.mdpi.com/1099-4300/20/11/875}.

\bibitem[{\citenamefont{Miller et~al.}(2021)\citenamefont{Miller, Mohammady,
  Perarnau-Llobet, and Guarnieri}}]{2021_miller_prl}
\bibinfo{author}{\bibfnamefont{H.~J.~D.} \bibnamefont{Miller}},
  \bibinfo{author}{\bibfnamefont{M.~H.} \bibnamefont{Mohammady}},
  \bibinfo{author}{\bibfnamefont{M.}~\bibnamefont{Perarnau-Llobet}},
  \bibnamefont{and}
  \bibinfo{author}{\bibfnamefont{G.}~\bibnamefont{Guarnieri}},
  \bibinfo{journal}{Phys. Rev. Lett.} \textbf{\bibinfo{volume}{126}},
  \bibinfo{pages}{210603} (\bibinfo{year}{2021}),
  \urlprefix\url{https://link.aps.org/doi/10.1103/PhysRevLett.126.210603}.

\end{thebibliography}
\end{document}


\preprint{APS/123-QED}

\title{Supplemental Material for ``A geometric bound on the efficiency of irreversible thermodynamic cycles"}

\author{Adam G. Frim}%
\affiliation{%
 Department of Physics, University of California, Berkeley, Berkeley, California, 94720
}%
\author{Michael R. DeWeese}
\affiliation{%
 Department of Physics, University of California, Berkeley, Berkeley, California, 94720
}%
\affiliation{%
Redwood Center For Theoretical Neuroscience,  University of California, Berkeley, Berkeley, California, 94720
}
\affiliation{%
Helen Wills Neuroscience Institute,University of California, Berkeley, Berkeley, California, 94720
}%

\date{\today}

\begin{abstract}

\end{abstract}

\maketitle


Throughout the Supplemental Material, all equations and figures are notated with an ``S" before the appropriate number. Any reference to equations or figures without an S refers to one in the main text.

\section{Isoperimetric bound on efficiency for arbitrary thermodynamic manifolds}

In the case when the curvature $K$ of the thermodynamic manifold $\mathcal{M}$ is of arbitrary and not necessarily constant sign, we must start with Eqs.~\eqref{eq:isoperimetric_bound_1} and \eqref{eq:isoperimetric_bound_2} of the main text:
\begin{align}
    \label{eq:S_isoperimetric_bound_1}
    &\mathcal{L}^2 \geq 4\pi \mathcal{A}-2\left[\iint_{\mathcal{C}} K^+ \right]\mathcal{A},\\
    \label{eq:S_isoperimetric_bound_2}
    &\mathcal{L}^2 \geq 4\pi \mathcal{A}-\bigg[\text{sup}_{\mathcal{C}} K \bigg]\mathcal{A}^2.
\end{align}
We no longer assume $K\leq 0$ such that we work with these equations directly. Following directly the same method as in the main text, we have
\begin{align}
    \label{eq:S_isoperimetric_work_bound_1}
    &\frac{\mathcal{L}^2}{\mathcal{W}} \geq 2\left(2\pi-\left[\iint_{\mathcal{C}}K^+ \right]\right) \frac{\mathcal{A}}{\mathcal{W}} = 2\left(2\pi-\left[\iint_{\mathcal{C}}K^+ \right]\right)\overline{\sqrt{g^{\mathcal{C}}}},\\
    \label{eq:S_isoperimetric_work_bound_2}
    &\frac{\mathcal{L}^2}{\mathcal{W}} \geq \left(4\pi-\bigg[\text{sup}_{\mathcal{C}} K\bigg]\mathcal{A}\right) \frac{\mathcal{A}}{\mathcal{W}} = \left(4\pi-\bigg[\text{sup}_{\mathcal{C}} K\bigg]\mathcal{A}\right) \overline{\sqrt{g^{\mathcal{C}}}}.
\end{align}
In both cases, we see that there is an effective correction to the bound in the main text. For $K>0$, Eqs.~\eqref{eq:S_isoperimetric_work_bound_1} and \eqref{eq:S_isoperimetric_work_bound_2} must be used in place of the bound given in the text. Importantly, for sufficiently large (positive) $K$ and/or $\mathcal{A}$, these bounds actually may yield $\mathcal{L}^2/\mathcal{W}\geq 0$, such that they provide no useful information. For $K\leq 0$, as before we still have $\mathcal{L}^2/\mathcal{W} \geq 4\pi \overline{\sqrt{g^{\mathcal{C}}}}$, though Eq.~\eqref{eq:S_isoperimetric_work_bound_2} gives further refinements and places stricter bounds (Eq.~\eqref{eq:S_isoperimetric_work_bound_1} degenerates to the simpler bound for $K\leq 0$). For the interested reader, we recommend the extensive review given by \cite{Osserman_1978} and references therein, specifically in \S 4, for a review of these forms of the classical isoperimetric inequalities, their geometric interpretation and implications, and their limitations.

\section{Isoperimetrically optimal Brownian protocols}
In the main text, we stated an isoperimetric bound and then found protocols that (nearly) saturated it for the specific case of the parametric harmonic oscillator in a variable heat bath. Here, we give a fuller account of our methods and demonstrate additional optimal protocols. In particular, starting with Eq.~(20) of the main text, we consider the following functional:
\begin{equation}
\label{eq:S_iso_functional}
    \mathcal{F} = \oint dt \left[\sqrt{g^\mathcal{C}} \tilde{P}\dot{\tilde{V}} + \xi \ \begin{pmatrix} \dot{\tilde{P}} \\ \dot{\tilde{V}} \end{pmatrix}
    \begin{pmatrix} \frac{\zeta}{2\tilde{P}} + \frac{2m\tilde{V}}{\zeta \tilde{P}} & \frac{m}{\zeta} \\
\frac{m}{\zeta} &
\frac{m\tilde{P}}{2\zeta \tilde{V}}
\end{pmatrix}
    \begin{pmatrix} \dot{\tilde{P}} \\ \dot{\tilde{V}} \end{pmatrix} \right].
\end{equation}
Standard arguments tell us setting the first variation of this functional to zero yields a constant geodesic curvature \cite{payne_isoperimetric_1967,Osserman_1978}. Explicitly, this constraint yields the Euler-Lagrange equations (in $\tilde{P}$ and $\tilde{V}$ variables):

\begin{align}
\label{eq:ELE_1}
&\xi  \tilde{P} \left[m \tilde{P} \left(\dot{\tilde{V}}^2-4 \tilde{V} \ddot{\tilde{V}}\right)-2 \tilde{V} \ddot{\tilde{P}} \left(\zeta ^2+4 m \tilde{V}\right)\right]+\xi  \tilde{V} \dot{\tilde{P}}^2 \left(\zeta ^2+4 m \tilde{V}\right)+\zeta  \tilde{P}^2 \sqrt{m \tilde{V}} \dot{\tilde{V}}=8 \xi  m \tilde{P} \tilde{V} \dot{\tilde{P}} \dot{\tilde{V}},\\
\label{eq:ELE_2}
&\tilde{P} \left[\dot{\tilde{P}} \left(2 \xi  m \tilde{V} \dot{\tilde{V}}+\zeta  \sqrt{m \tilde{V}^3}\right)+4 \xi  m \tilde{V}^2 \ddot{\tilde{P}}\right]=\xi  m \left[4 \tilde{V}^2 \dot{\tilde{P}}^2+\tilde{P}^2 \left(\dot{\tilde{V}}^2-2 \tilde{V} \ddot{\tilde{V}}\right)\right].
\end{align}

\begin{figure}[t]
    \centering
    \includegraphics[width=\linewidth]{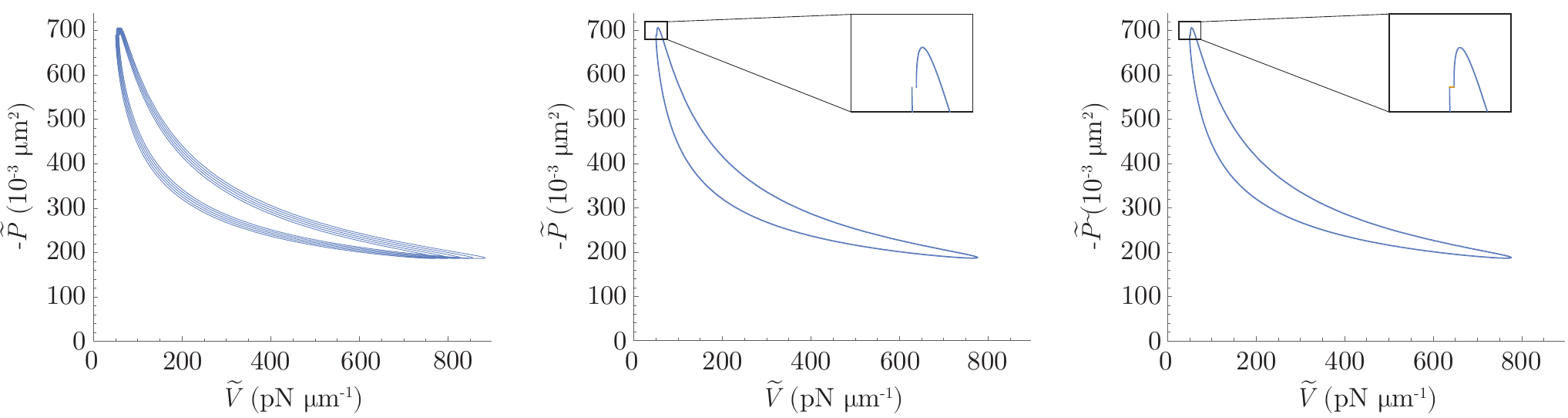}
    \caption{Characteristic example of optimal cycle construction. Left panel: A candidate non-closed curve that solves Eqs.~\eqref{eq:ELE_1} and \eqref{eq:ELE_2} for given initial conditions integrated over a set time domain. Middle panel: The candidate curve is truncated at the closest point tangent to the initial condition after a single ``period." A magnified section of the curve at this truncation point is shown in the inset demonstrating the non-closed nature of the curve. Left Panel: A cycle is formed by closing the curve with a straight line path (shown in orange).}
    \label{fig:figS1}
\end{figure}

Solving these equations therefore yields curves that have constant geodesic curvature and are therefore candidates to be isoperimetrically optimal. Over a given domain, the \textit{isoperimetric profile} gives the (perhaps degenerate set of) minimizer(s) of perimeter for a fixed area, and can be shown to be closed and smooth; furthermore, given certain demands on connectedness and compactness of a domain \cite{Osserman_1978}, it can be shown to exist. Nevertheless, its existence is not equivalent to its construction, and finding exactly isoperimetrically optimal shapes is a significant undertaking in the mathematics literature, especially for surfaces of non-constant, non-symmetric curvature.

To avoid this significant difficulty, we bypass the problem of finding exactly closed curves of constant geodesic curvature and instead simply look for closed, not-necessarily-smooth curves that maintain constant curvature everywhere excluding small regions. Finding such cycles is a far simpler task and they still perform optimally relative to the bound, Eq. (17) of the main text. In particular, Starting with Eqs. \eqref{eq:ELE_1} and \eqref{eq:ELE_2}, we numerically integrate for the given set of initial conditions of $\tilde{P}(0), \tilde{V}(0), \dot{\tilde{P}}(0),$ and $\dot{\tilde{V}}(0)$, and 
a given value of $\xi$. As noted in the main text, if we are only interested in the resultant curve rather than its particular parameterization, we could instead parameterize $\tilde{P}$ as a function of $\tilde{V}$, such that one of these initial condition degrees of freedom is redundant so optimal shapes under this construction are specified by a four-parameter family. Like the variational optimal isoperimentric construction of a circle, the Lagrange multiplier $\xi$ enforcing constant curve energy $\oint dt (\dot{\tilde{P}}, \dot{\tilde{V}})\cdot\mathbf{g}\cdot  (\dot{\tilde{P}}, \dot{\tilde{V}})$ correlates to a diameter-like quantity that specifies the size of the resultant cycle.

Generically, and typically, the cycles we find are not closed and instead result in cycloids in space. However, seeing as we are interested only in closed cycles for engine operation, we find the first point on the curve whose tangent vector is tangent to that initial condition, empirically these tend to be quite close, and connect them by a straight line. A characteristic example of this process is demonstrated in Fig.~\ref{fig:figS1} for a specific choice of initial conditions. Generally speaking, this small correction to close the cycle amounts to an added value of thermodynamic length that is significantly shorter than that of the rest of the cycle (and therefore energetically cheap relative to the total cycle), but we still do recognize the lack of true optimality of these cycles. Nevertheless, compared to the bound, they perform quite well, as can be seen in Fig. 3 of the main text. Empirically, we find these cycles can approach the bound to within tens of thousandths of a percent and are never more than a handful of percentages above it for the parameter values we explore.
\begin{figure}[t]
    \centering
    \includegraphics[width=\linewidth]{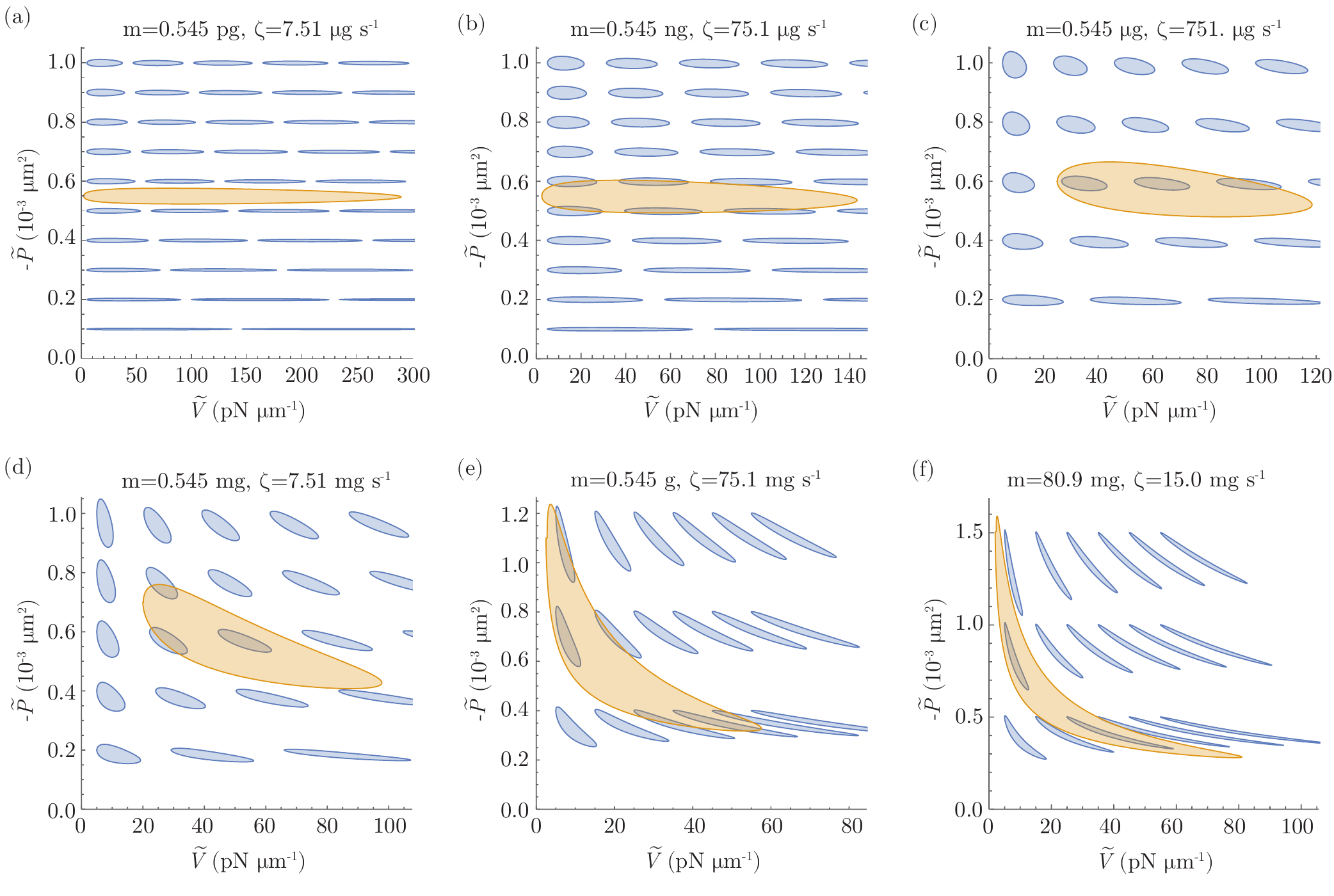}
    \caption{Plots of optimally constructed cycles for a variety of material parameters and in a variety of regions of Clausius space. The specific values of $m$ and $\zeta$ relevant to each plot are displayed above them. All smaller blue plots contain the same Euclidean area (and therefore quasistatic work) of $6$ pN $\mu$m and the larger orange plots contain $120$ pN $\mu$m. Note, (a) and (f) are reproduced from Fig.~\ref{fig:fig2}(a) and (b), respectively, to be compared with cycles for other choices of $m$ and $\zeta$.}
    \label{fig:figS2}
\end{figure}

In Fig.~\ref{fig:figS2}, we plot such isoperimetrically optimized curves for a variety of values of $m$ and $\zeta$, which impact the metric and curvature of the underlying manifold, $K = 1/(\zeta \tilde{P})$. The curves themselves can further be understood by studying the metric
\[g_{\mu\nu}^{\mathcal{C}} =-\begin{pmatrix} \frac{\zeta}{2\tilde{P}} + \frac{2m\tilde{V}}{\zeta \tilde{P}} & \frac{m}{\zeta} \\
\frac{m}{\zeta} &
\frac{m\tilde{P}}{2\zeta \tilde{V}}
\end{pmatrix},
\]
where the ratio of various terms, including the fixed ratio $m/\zeta$ set by material parameters, show the competition between various directions of minimal length. Considering these ratios (which \emph{do} have dimensions) can give intuition into why a cycle might favor being heavily horizontal, as in Fig.~\ref{fig:fig1} of the main text and Figs \ref{fig:figS2}(a)-(d) of this supplement. In all plots, blue cycles enclose an quasistatic work (given by their enclosed Euclidean area in these units) of $6$ pN $\mu$m and orange cycles enclose 120 pN $\mu$m, such that we can observe the resultant cycles over multiple scales. 

We observe that for smaller values of $m/\zeta$, the cycles are overwhelmingly horizontal (in the plotted units). As noted previously, evaluating sample values of the metric for this case, this makes sense as the cost to change horizontally is far cheaper than to change vertically. As $m/\zeta$ increases, however, the optimal curves become more curved and seemingly follow more or less along adiabats. In a previous work, we showed adiabats of this model are exactly geodesic \cite{2022_PRE_Frim}, so this empirical finding makes intuitive sense as these are the least costly means of connecting any given points in Clausius space. More generally, characterizing the full set of optimal curves of this model as a function of $m$ and $\zeta$ is a difficult but worthwhile analytical and numerical task which we will leave to future work.

\section{Classical Brownian protocols}
In Fig.~\ref{fig:fig3} of the main text, we compared the optimally constructed curves detailed previously to a variety of cycles. Here, we review notable classical thermodynamic processes and cycles and how they are manifested in Brownian colloidal systems, following \cite{2011_NatPhys_Blickle,2015_PRL_Martinez,2016_NatPhys_Martinez,2017_SM_Martinez}.
\subsection{Thermodynamic Processes}
\subsubsection{Isothermal Processes}
For Brownian systems, isothermal processes retain their usual meaning in that temperature is maintained throughout the process. However, given the stochastic and nonequilibrium nature of the working system, a notion of temperature may be ill-defined for the system itself and so the bath temperature $T$ is maintained while other control variables are allowed to vary, specifically the trap stiffness $k$ for the working substance considered in the main text.

In Clausius space, we have $\tilde{V} \equiv k$ and $\tilde{P}\equiv -k_BT/(2k)$ such that $\tilde{P}\tilde {V} =-k_B T/2$ and the equation of state of this system mirrors a (virtual) ideal gas with $N=-1/2$, and an isothermal process thus corresponds to 
\begin{equation}
T  = \text{const.} \iff (\tilde{P}\tilde{V})_{\text{isothermal}} = \text{const.}
\end{equation}
as in the case of an ideal gas.
\subsubsection{Adiabatic Processes}
In classical thermodynamics, an adiabatic process is one that occurs with no heat exchange, and is therefore exactly isentropic for reversible processes. Although these use interchangeably, there can be irreversible adiabatic processes that are therefore \textit{not} isentropic. In the Brownian context, adiabatic processes are defined as those that maintain the system entropy were the process occurring quasistatically. 

In the case of the parametric harmonic oscillator, the Boltzmann distribution of states is given
\begin{equation}
\rho_{\text{Boltzmann}} = \frac{1}{Z} e^{-\beta (\frac{p^2}{2m}+\frac{1}{2}kx^2)},
\end{equation}
where $\beta \equiv (k_BT)^{-1}$ is the inverse temperature, $p$ is the particle momentum, $x$ is the particle position, and the partition function is given
\begin{equation}
Z = 2\pi k_BT \sqrt{\frac{m}{k}}.
\end{equation}
The Helmholtz free energy is thus
\begin{equation}
F = -k_B T \log Z = -k_BT \log \left( 2\pi k_BT \sqrt{\frac{m}{k}}\right),
\end{equation}
such that, finally, the equilibrium system entropy for fixed $k$ and $T$ is 
\begin{equation}
S = -\frac{\partial F}{\partial T} = k_B \left[1+\log \left( 2\pi k_BT \sqrt{\frac{m}{k}}\right)\right].
\end{equation}
Therefore, a Brownian adiabatic process is defined:
\begin{equation}
S = \text{const.} \iff \left(\frac{T^2}{k}\right)_\text{adiabitc} = \text{const.} \iff \left(\tilde{P}\tilde{V}^{1/2}\right)_\text{adiabatic} = \text{const.} 
\end{equation}

Note for overdamped Brownian systems, although the definition of adiabatic processes remains the same, the system's inertial degrees of freedom are ignored such that $S$ becomes a function of $T/k$ directly and adiabatic processes are those that maintain this ratio. To avoid confusion, these processes are instead typically denoted as pseudo-adiabatic \cite{2015_PRL_Martinez}.

\subsubsection{Isochoric Processes}
Like isothermal processes, isochoric processes retain their standard definition in that the main control variable is maintained. In control space, this manifests as $k = \text{const.}$ such that
\begin{equation}
(k)_\text{isochoric} = \text{const.} \iff (\tilde{V})_\text{isochoric} = \text{const.} 
\end{equation}

\subsubsection{Geodesic Processes}
Geodesics are the shortest paths connecting two given points on a manifold and generalize straight-line paths to non-Euclidean spaces (e.g. great circles are the geodesics on the surface of a sphere). Recently, geodesics have been suggested as a means to construct thermodynamic processes \cite{2012_PRE_zulkowski,2022_PRE_Frim}. Specifically, the geodesics of the thermodynamic manifold connecting two given points in control space are calculated and the resulting path in control space is followed as the process. Given their definition, these processes correspond to those of shortest thermodynamic length between any two points in control space and are therefore the least dissipative between those two points. Generically, such paths must be found numerically, as is the case for this model system \cite{2022_PRE_Frim}.

\subsection{Thermodynamic Cycles}

\begin{figure}[t]
    \centering
    \includegraphics[width=\linewidth]{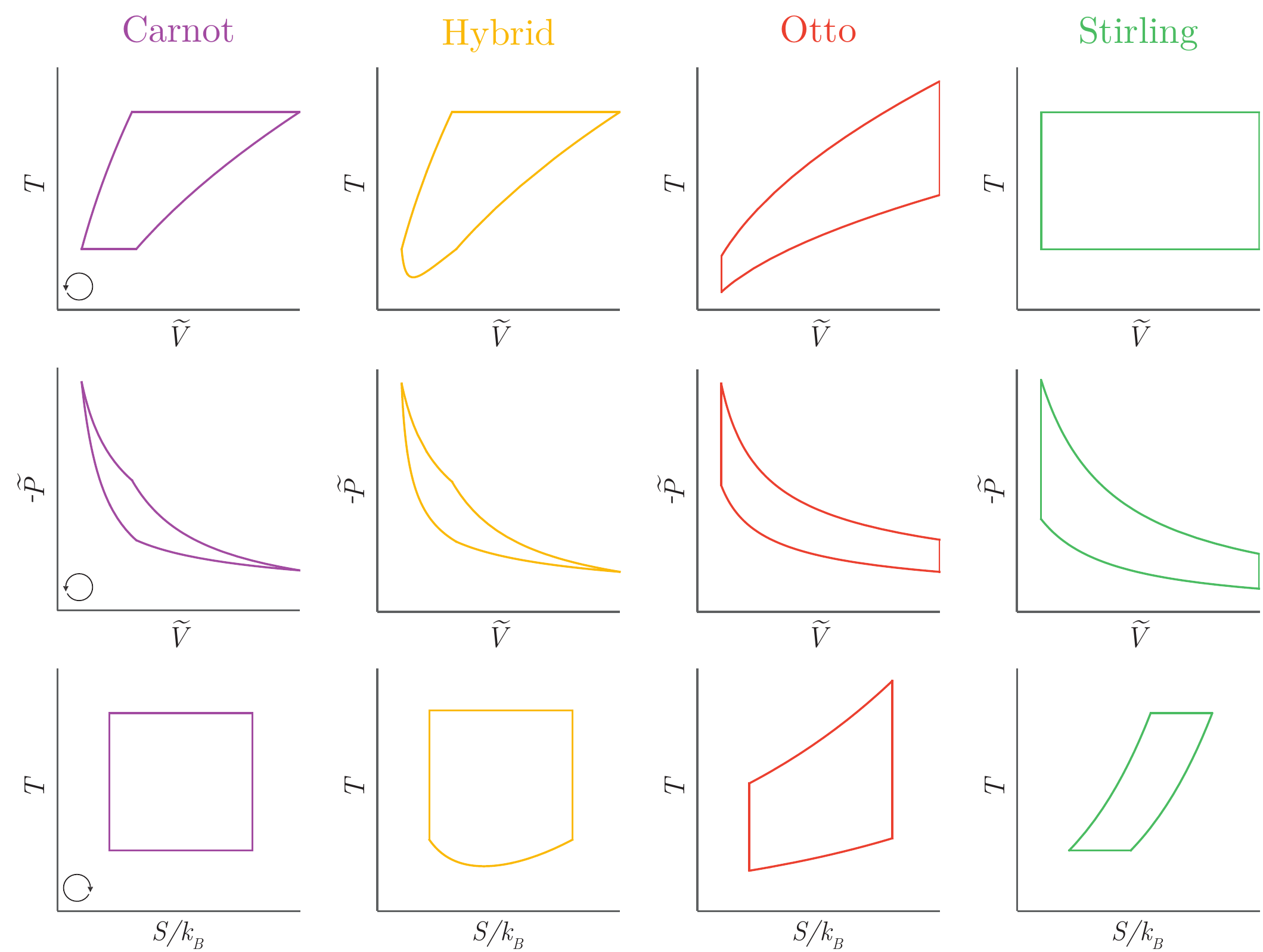}
    \caption{Quasistatic profiles of classical thermodynamic cycles in various spaces. Carnot cycle (far left, purple), Hybrid cycle (middle left, yellow), Otto cycle (middle right, red), and Stirling cycle (far right, green) shown in control ($T-\tilde{V}$) space, Clausius ($\tilde{P}-\tilde{V}$) space and $T-S$ space. Arrows in Carnot cycle profiles denote orientation for positive work output: the same orientations produce positive work output for all other cycles. }
        \label{fig:figS3}
\end{figure}

\subsubsection{Brownian Carnot cycle}
Following the usual definition of the Carnot cycle, a Brownian Carnot cycle consists of alternating isothermal and adiabatic strokes. A schematic quasistatic profile in control space, Clausius space and $T-S$ space are shown on the far left of Fig.~\ref{fig:figS3} in purple.

\subsubsection{Brownian Hybrid cycle}
Hybrid cycles were introduced in \cite{2022_PRE_Frim} and consist of a Carnot cycle with slight modification. Namely, heat is extracted from a high temperature bath by means of an isothermal stroke and connected to a low temperature bath by means of adiabatic strokes. However, in lieu of an exhaust isothermal stroke at a low temperature bath, these two corners of the usual Carnot cycle are connected by means of a geodesic stroke, such that the cycle is a hybrid between a Carnot cycle and a geodesic cycle lending the cycle its name. This geodesic stroke is, by definition, less dissipative than the corresponding isothermal stroke connecting these two corners, though proves to provide more output work. As a result, the Hybrid cycle is less dissipative, more efficient, and more powerful for a given cycle duration than its corresponding Carnot cycle. A schematic quasistatic profile in control space, Clausius space and $T-S$ space are shown on the middle left of Fig.~\ref{fig:figS3} in yellow.

\subsubsection{Brownian Otto cycle}
Following the usual definition of the Otto cycle, a Brownian Otto cycle consists of alternating isochoric and adiabatic strokes. A schematic quasistatic profile in control space, Clausius space and $T-S$ space are shown on the middle right of Fig.~\ref{fig:figS3} in red.

\subsubsection{Brownian Stirling cycle}
Following the usual definition of a Stirling cycle, a Brownian Stirling cycle consists of alternating isothermal and isochoric strokes. A schematic quasistatic profile in control space, Clausius space and $T-S$ space are shown on the far right of Fig.~\ref{fig:figS3} in green.

\section{Construction of Fig.~3}
In this section, we give further details on how Fig.~3 of the main text is constructed. First, we detail the construction of the ``named" cycles. For each named cycle, we randomly selected the corners of the cycle in Clausius space to construct the sufficient number of cycles. In the case of the Carnot cycle, this amounted to randomly selecting the maximum and minimum temperature and entropy; for the Stirling, temperature and volume; for the Otto, volume and entropy; and for the Hybrid, the same corners were considered as a randomly constructed Carnot. To plot over a sufficient region of space, these cycles were sampled such that a significant number would appear for a wide range of $\overline{\sqrt{g^\mathcal{C}}}$, amounting to a wide range of average values for $1/\sqrt{\tilde{V}}$.

For the randomly constructed cycles, denoted by the gray X's in the figure, cycles are specified by $-\tilde{P}(t)=f_1(\sin(x))+f_2(\cos(x))$ and $\tilde{V}(t)=f_3(\sin(x))+f_4(\cos(x))$ where the functions are of the form $f_i(x) = c_{i0}+c_{i1}x+c_{i3}x^3+c_{i5}x^5$ with random coefficients $c_{ij}$, and where $x$ is evaluated over the full period $[0,2\pi]$. Given that thermodynamic space for this model system is only well-defined for $\tilde{V}>0$ and $\tilde{P}<0$, there are nevertheless constraints on $c_{ij}$ such that $f_i$ are all positive. First, we choose $c_{10}+c_{20}$ a randomly selected positive number over a uniform distribution whereas $c_{30}+c_{40}$ is selected from a Pareto distribution with minimum value 0.02 and shape parameter 0.5 such that again cycles are sampled from a large range of of average values for $1/\sqrt{\tilde{V}}$ (roughly uniform over the range $[0,100])$. From there $c_{i1}\sim \mathcal{U}_{[-c_{i0}/2,c_{i0}/2]}, c_{i3}\sim \mathcal{U}_{[-(c_{i0}/2-|c_{i1}|),(c_{i0}/2-|c_{i1})]},$ and $c_{i5}\sim \mathcal{U}_{[-(c_{i0}/2-|c_{i1}|-|c_{i3}|),(c_{i0}/2-|c_{i1}-|c_{i3})|]}$ where $\mathcal{U}_{[a,b]}$ is the continuous uniform distribution over the range $[a,b]$. By these selections, all $f_{i}$ are guaranteed positive functions such that the cycles are well-defined, randomly constructed, smooth and closed thermodynamic cycles that sample a large region of Clausius space. Altogether, 30000 such cycles were constructed, though a significant number have values of $\mathcal{L}^2/\mathcal{W}$ that are outside the plotted range of Fig.~3.

Finally, the optimally constructed cycles were selected by choosing 500 sets of values for $\tilde{P}(0), \tilde{V}(0), \dot{\tilde{P}}(0), \dot{\tilde{V}}(0)$, and $\xi$ and solving Eqs.~\eqref{eq:ELE_1} and \eqref{eq:ELE_2} numerically. Values were sampled over a large range of average values of $1/\sqrt{\tilde{V}}$ to demonstrate their performance against the bound over a wide region of Clausius space.

For all such constructed cycles, the thermodynamic lengths, areas, and quasistatic work values (i.e. Euclidean areas) are calculated (these are all purely geometric quantities), and used to construct the figure. 

\section{Isoperimetric inequalities with density}
Manifolds with density have become of interest recently in the mathematics literature \cite{morgan_2}. These mathematical structures, which we will denote as $\mathcal{M}_\psi$ consist of a base manifold $\mathcal{M}_0$ with usual definitions of volume $\mathscr{V}$ and area $\mathscr{A}$ (or area and length in the case of two dimensions) but further equipped with a defined set of densities $\{\psi\}$ such that the volume $\mathscr{V}_\psi$ and area $\mathscr{A}_\psi$ become 
\[\begin{split}
    &d\mathscr{V} = \psi_\mathscr{V}d\mathscr{V}_0,\\
    &  d\mathscr{A} = \psi_\mathscr{A} d\mathscr{A}_0.
\end{split} \]
Generalizing various results, notably isoperimetric inequalities, for standard manifolds to manifolds with density  has been on ongoing endeavor with success for certain specific cases, most notably the Gauss space: Euclidean space with a normalized Gaussian density centered at the origin for both area and volume. 

Of interest to us is the recognition that, in Clausius space, $d\mathcal{W} = d\mathcal{X}_\lambda d\lambda $ whereas $d\mathcal{A} = \sqrt{g^{\mathcal{C}}}d\mathcal{X}_\lambda d\lambda $. In the main text, we leveraged the ratio of these two quantities to place on bound on $\mathcal{L}^2/\mathcal{W}$. However, recognizing \begin{equation}
   d \mathcal{W} = \frac{1}{\sqrt{g^{\mathcal{C}}}}d\mathcal{A},
\end{equation}
suggests that if we consider the manifold with density consisting of
\begin{enumerate}
    \item A base manifold of Clausius space
    \item A trivial (i.e. unit) length density
    \item An area density consisting of $1/\sqrt{g^\mathcal{C}}$
\end{enumerate}
Then finding isoperimetric inequalities and isoperimetric curves on this manifold with density would yield general bounds on $\mathcal{L}^2/\mathcal{W}$ and, in turn, efficiency. Solving such a problem generally, or even in a specific case, is a highly nontrivial task; nevertheless, the significant impact such results would have merits their investigation.